\documentclass[prd,showpacs,superscriptaddress,nofootinbib,floatfix,11pt]{revtex4-2}
\usepackage{setspace}
\usepackage[utf8]{inputenc}
\usepackage{float}
\usepackage{amsmath,amssymb,amsfonts,bm}
\usepackage{graphicx}
\usepackage{multirow}
\usepackage{xcolor}
\usepackage{slashed}
\allowdisplaybreaks 
\usepackage[
colorlinks=true,
linkcolor=blue,
breaklinks=true,
citecolor=blue]{hyperref}

\newcommand{\dsds}{D_s^+D_s^-}
\newcommand{\vsl}{\slashed{v}}
\newcommand{\Tr}{\text{Tr}}

\newcommand{\itp}{\affiliation{CAS Key Laboratory of Theoretical Physics, Institute of Theoretical Physics,\\ Chinese Academy of Sciences, Beijing 100190, China}}

\newcommand{\ucas}{\affiliation{School of Physical Sciences, University of Chinese Academy of Sciences, Beijing 100049, China}}

\newcommand{\ific}{\affiliation{Instituto de F\'isica Corpuscular (centro mixto CSIC-UV),
Institutos de Investigaci\'on de Paterna, \\ Apartado 22085, 46071 Valencia, Spain
}}

\newcommand{\uestc}{\affiliation{School of Physics, University of Electronic Science and Technology of China, \\ Chengdu 611731, China}}

\begin{document}

\title{\boldmath Establishing the heavy quark spin and light flavor molecular multiplets of the $X(3872)$, $Z_c(3900)$ and $X(3960)$}

\author{Teng~Ji}\email{jiteng@itp.ac.cn}
\itp \ucas

\author{Xiang-Kun~Dong}\email{dongxiangkun@itp.ac.cn}
\itp \ucas

\author{Miguel~Albaladejo}\email{Miguel.Albaladejo@ific.uv.es}
\ific

\author{Meng-Lin~Du}\email{du.menglin@ific.uv.es}
\ific\uestc

\author{Feng-Kun~Guo}\email{fkguo@itp.ac.cn}\itp \ucas

\author{Juan~Nieves}\email{jmnieves@ific.uv.es}
\ific



\begin{abstract} 
  Recently, the LHCb Collaboration reported a near-threshold enhancement, $X(3960)$, in the $D_s^+D_s^-$ invariant mass distribution. We show that the data can be well described by either a bound  or a virtual state below the $D_s^+D_s^-$ threshold. The mass given by the pole position is $(3928\pm3)$~MeV. 
  Using this mass and the existing information on the $X(3872)$ and $Z_c(3900)$ resonances, a complete spectrum of the $S$-wave hadronic molecules formed by a pair of ground state charmed and anticharmed mesons is established. Thus, pole positions of the partners of the $X(3872)$, $Z_c(3900)$ and the newly observed $D_s^+D_s^-$ state are predicted. Calculations have been carried out at the leading order of nonrelativistic effective field theory and considering both heavy quark spin  and light flavor SU(3) symmetries, though conservative errors from the breaking of these symmetries are provided.
\end{abstract}

\maketitle

\section{Introduction}

The $X(3872)$~\cite{Belle:2003nnu}, also known as $\chi_{c1}(3872)$, and the $Z_c(3900)$~\cite{BESIII:2013ris,Belle:2013yex} have been proposed as candidates of isoscalar and isovector $D\bar D^*$ hadronic molecules, respectively, since long~\cite{Barnes:2003vb,Voloshin:2003nt,Swanson:2003tb,Tornqvist:2004qy, Swanson:2004pp, Suzuki:2005ha,AlFiky:2005jd, Gamermann:2006nm,Hanhart:2007yq, Fleming:2007rp,Braaten:2007dw, Thomas:2008ja,Fleming:2008yn,Liu:2008fh, Liu:2009qhy,Gamermann:2009fv, Gamermann:2009uq,Bignamini:2009sk, Nieves:2011zz, Nieves:2012tt,Guo:2013zbw,Voloshin:2013dpa,Wang:2013vex, Hidalgo-Duque:2012rqv,Prelovsek:2013cra,Prelovsek:2013cra,Guo:2013sya,Hidalgo-Duque:2013pva,Guo:2014iya,Guo:2014hqa,Guo:2014taa,Swanson:2014tra,Albaladejo:2015dsa,Albaladejo:2015lob,Baru:2015nea,Szczepaniak:2015eza,Albaladejo:2016jsg,Chen:2016qju,Cincioglu:2016fkm,Albaladejo:2017blx,Guo:2017jvc,Guo:2019qcn,Voloshin:2019ivc,Yang:2020nrt,Zhang:2020mpi,Dong:2020hxe,Brambilla:2019esw,Dong:2021bvy, Baru:2021ddn, Du:2022jjv}.
They are expected to have heavy quark spin and light flavor SU(3) siblings~\cite{Hidalgo-Duque:2012rqv}, with the observed $Z_c(4020)$~\cite{BESIII:2013ouc} and $Z_{cs}(3985)$~\cite{BESIII:2020qkh,LHCb:2021uow} structures being the candidates for the $Z_c(3900)$ partners.

Very recently, the LHCb Collaboration announced the observation of a new resonant structure $X(3960)$ in the $\dsds$ invariant mass distribution of the $B^{+} \rightarrow \dsds K^{+}$ decay~\cite{LHCb:2022NewObservations}. The peak structure is just above the $\dsds$ threshold with a statistical significance larger than $12\sigma$. The mass and width reported by LHCb, from an analysis using a Breit-Wigner (BW) parametrization, are
\begin{equation}
    M=(3955 \pm 6 \pm 11) ~\mathrm{MeV},  \quad \Gamma=(48 \pm 17 \pm 10) ~\mathrm{MeV}, \label{eq:lhcb}
\end{equation}
respectively, with $J^{P C}=0^{++}$ favored quantum numbers.
This resonance couples to the $\dsds$ in an $S$ wave, and is an excellent candidate for a $\dsds$ hadronic molecule, which has been predicted in various theoretical calculations~\cite{Hidalgo-Duque:2012rqv,Li:2015iga,Prelovsek:2020eiw,Meng:2020cbk,Dong:2021juy}. 
In particular, a recent lattice work predicted a shallow $\dsds$ bound state below the threshold with a binding energy $2m_{D_s} - M = 6.2^{+2.0}_{-3.8} ~\mathrm{MeV}$~\cite{Prelovsek:2020eiw}, and the vector-meson-dominance model in Ref.~\cite{Dong:2021juy} predicted a shallow virtual state with a virtual energy roughly in the range of $[4.7,35.5]$~MeV.
The difference between a bound and a virtual state is that the former is a pole of the scattering amplitude below threshold on the real axis of the first Riemann sheet (RS) of the complex energy plane, while the latter is on the second RS.
{  The bound state found in the lattice calculation~\cite{Prelovsek:2020eiw} becomes a narrow resonance when the $D\bar D$ channel is coupled to the $D_s\bar D_s$; however, it still couples predominantly to the $D_s\bar D_s$ and owes its origin to the interactions of the $D_s\bar D_s$ pair. In this sense and in an abuse of language, we will refer to such a pole as a ``bound state'',  even if it becomes metastable  due to the opening of the lower channel which weakly couples to the pole. Such a notation is also widely used in the literature; e.g., the positronium is ubiquitously called an $e^+e^-$ bound state although it decays into photons.}

The LHCb measurements of the $\dsds$ near-threshold structure~\cite{LHCb:2022NewObservations}, the mass of the $X(3872)$~\cite{LHCb:2020xds}, the isospin breaking in its decays~\cite{LHCb:2022bly}, and the BESIII measurements of the $Z_c(3900)$~\cite{BESIII:2015pqw,BESIII:2017bua} and $Z_{cs}(3985)$~\cite{BESIII:2020qkh} allow us to make a prediction of the full spectrum of the $S$-wave hadronic molecules formed by a pair of ground state charmed and anticharmed mesons, at leading order (LO) of the nonrelativistic effective field theory (NREFT)~\cite{AlFiky:2005jd,Mehen:2011yh,Nieves:2012tt,Hidalgo-Duque:2012rqv,Guo:2013sya,Guo:2017jvc}. 

\section{NREFT}

Since the charm quark mass, $m_c \sim 1.5$~GeV, is much larger than the nonperturbative scale of Quantum Chromodynamics (QCD), $\Lambda_{\rm QCD}\sim 0.3$~GeV, the interaction depending on the charm quark spin inside hadrons is suppressed, compared with the spin-independent one, by a factor of  $\mathcal{O}(\Lambda_{\rm QCD}/m_c) \sim 0.2$. Consequently, there is an approximate heavy quark spin symmetry (HQSS), and hadrons containing a charm (heavy)  quark fall into  spin multiplets characterized by the parity of the hadrons and the total angular momentum of the light degrees of freedom $j_\ell^P$, including the spin of the light quarks and gluons and the orbital angular momentum.  The ground state pseudoscalar and vector charmed mesons $D$ and $D^*$ belong to the multiplet  with $j_\ell^P = 1/2^-$, denoted in this work as $H$.

In the energy region near the thresholds of a pair of $j_\ell^P = 1/2^-$ charmed and anticharmed mesons, the approximate HQSS is also manifested in the interaction between the two hadrons in the pair. At LO of the heavy quark and nonrelativistic expansion, the $S$-wave interaction between a pair of charmed and anticharmed mesons can be parametrized into constant contact terms, which are then inserted into the Lippmann-Schwinger equation (LSE) to solve for poles (see, e.g., Ref.~\cite{Hidalgo-Duque:2012rqv} and the next section). These poles would correspond to hadronic molecules formed by the meson pairs.
The total angular momentum of the light degrees of freedom for the $H\bar H$ pair can have $n_\ell=2$ possible values: $0$ and $1$. 
If we further consider the light flavor SU(3) symmetry, all $H\bar H$ pairs form $n_f=2$ irreducible representations. This follows from the reduction $[\bar 3] \otimes [3] = [1] \oplus [8]$, where $[\bar 3]$ and $[3]$ (SU(3) antitriplet and triplet, respectively) stand for   the light quark content of charmed and anticharmed mesons.
Therefore, at LO, there are only $n_\ell\times n_f=4$ independent constant contact terms, which are denoted here as $\mathcal C_{0a}, \mathcal C_{0b}, \mathcal C_{1a}$ and $\mathcal C_{1b}$.
For more details we refer to Ref.~\cite{Hidalgo-Duque:2012rqv}, though we will explicitly construct below the LO Lagrangian. In  brief, 
isoscalar molecules ($I=0, S=0$) without hidden strangeness can be described with $\mathcal C_{0a}$ and $\mathcal C_{0b}$. Isospinor
($I=1/2, S=\pm 1$) and isovector ($I=1, S=0$) octet states are in turn determined by $\mathcal C_{1a}$ and $\mathcal C_{1b}$. Finally, for molecular states with hidden strangeness, the contact range interactions are the average of the isoscalar and isovector ones.

We use the matrix field $H^{(Q)}$ [$H^{(\bar Q)}$] to
describe the combined SU(3) antitriplet [triplet] of pseudoscalar heavy mesons
$P^{(Q)}_a=(Q\bar u,Q\bar d, Q\bar s)$ [$P^{(\bar Q)a}=(u \bar Q,d \bar Q, s\bar Q )$] fields, with $a$ the light-flavor index, and
their vector HQSS partners $P^{*(Q)}_a$ [$P^{*(\bar Q)a}$]~\cite{Grinstein:1992qt},
\begin{eqnarray}
H_a^{(Q)} &=& \frac{1+\vsl}2 \left (P_{a\mu}^{* (Q)}\gamma^\mu -
P_a^{(Q)}\gamma_5 \right), \qquad\quad v\cdot P_{a}^{* (Q)} = 0,  \nonumber \\
H^{(\bar Q)a} &=&  \left (P_{\mu}^{* (\bar Q)a}\gamma^\mu -
P^{(\bar Q)a}\gamma_5 \right) \frac{1-\vsl}2 , \qquad v\cdot P^{*
  (\bar Q)a} = 0.
\end{eqnarray}
The matrix field $H^{(Q)}$ [$H^{(\bar Q)}$] annihilates $P^{(Q)}$ [$P^{(\bar Q)}$]
and $P^{*(Q)}$ [$\bar P^{*(\bar Q)}$] mesons with a definite velocity $v$. The definition for $H^{(\bar Q)a}$ also
specifies our convention for charge conjugation, which is $\mathcal{C}P_a^{(Q)}
\mathcal{C}^{-1} = P^{(\bar Q) a} $ and $\mathcal{C}P_{a\mu}^{*(Q)}\mathcal{C}^{-1}
= -P_\mu^{*(\bar Q) a}$.\footnote{It follows from
\begin{equation}
\mathcal{C}H_a^{(Q)} \mathcal{C}^{-1} = c\, H^{(\bar Q)aT}\, c^{-1}, \qquad
\mathcal{C}\bar H^{(Q)a}\mathcal{C}^{-1} = c\, \bar H_a^{(\bar Q)T}\, c^{-1},
\end{equation}
with $c$ the Dirac space charge conjugation matrix satisfying $c\gamma_\mu
c^{-1}=-\gamma_\mu^T$.}

The field
$H_a^{(Q)}$ [$H^{(\bar Q)a}$] transforms as a $(2,\bar 3)$ [$(\bar
  2,3)$] under the heavy-quark spin $\otimes $ SU(3), with the latter referring to the light-flavor symmetry~\cite{Grinstein:1992qt},
\begin{equation}
H_a^{(Q)} \to S \left( H^{(Q)} U ^\dagger\right)_a, \qquad
H^{(\bar Q) a} \to \left(U  H^{(\bar Q)}\right)^a S^\dagger,
\end{equation}
with $S$ and $U$ the transformation matrices acting on the heavy quark spin and SU(3) flavor spaces, respectively.
Their hermitian conjugate fields are defined by:
\begin{equation}
\bar H^{(Q)a} =\gamma^0 H_a^{(Q)\dagger} \gamma^0, \qquad
\bar H_a^{(\bar Q)} =\gamma^0 \bar H^{(\bar Q)a\dagger} \gamma^0 ,
\label{eq:Hdag}
\end{equation}
and transform as~\cite{Grinstein:1992qt}:
\begin{equation}
\bar H^{(Q)a} \to  \left( U \bar H^{(Q)} \right)^a S^\dagger , \qquad
\bar H^{(\bar Q)}_a \to S\left(\bar H^{(\bar Q)} U^\dagger \right)_a .
\end{equation}

At very low energies, the
interaction between a pair of heavy and anti-heavy mesons can be accurately
described just in terms of a contact-range potential.
The LO Lagrangian respecting HQSS reads~\cite{AlFiky:2005jd}
\begin{eqnarray}
\label{eq:LaLO}
\mathcal{L}_{4H} & = & \frac{1}{4}\,\Tr\left[\bar{H}^{(Q)a}{H}_b^{(Q)} \gamma_{\mu}
\right] \Tr\left[{H}^{(\bar{Q})c} \bar{H}^{(\bar{Q})}_d \gamma^{\mu} \right] \left( F_{A}^{}\, \delta_a^{\,b} \delta_{c}^{\,d} + F_{A}^{\lambda}\, \vec\lambda_a^{\,b} \cdot \vec\lambda_c^{\,d} \right)
\nonumber\\ 
\nonumber\\
&&+ \frac{1}{4}\,\Tr\left[\bar{H}^{(Q)a}{H}_b^{(Q)} \gamma_{\mu}\gamma_5
\right] \Tr\left[{H}^{(\bar{Q})c} \bar{H}^{(\bar{Q})}_d \gamma^{\mu}\gamma_5
\right] \left( F_{B}^{}\, \delta_a^{\,b} \delta_{c}^{\,d} + F_{B}^{\lambda}\, \vec\lambda_a^{\,b} \cdot \vec\lambda_c^{\,d} \right),
\end{eqnarray}
with $\vec\lambda$ the eight Gell-Mann
matrices in the SU(3) flavor space, and $F_{A,B}^{(\lambda)}$ light-flavor independent low-energy constants (LECs). Here, $\Tr$ takes trace in the  spinor space, and $\vec\lambda\cdot \vec\lambda$ sums over all Gell-Mann matrices. Note that in our normalization the heavy or anti-heavy meson fields, $H^{(Q)}$ or $H^{(\bar Q)}$, have dimensions of
energy$^{3/2}$ (see \cite{Manohar:2000dt} for details). This is because
we use a nonrelativistic normalization for the heavy mesons, which
differs from the traditional relativistic one by a factor
  $\sqrt{2M_H}$. 
The four LECs that appear above  are rewritten into
$\mathcal C_{0a}$, $\mathcal C_{0b}$ and $\mathcal C_{1a}$, $\mathcal C_{1b}$, which stand for the LECs in the isospin
$I = 0$ and $I = 1$ channels, respectively. The
relations read
\begin{eqnarray}
\mathcal C_{0a} &=& F_{A} + \frac{ 10 F_{A}^{\lambda}}{3}, \qquad
\mathcal C_{1a} = F_{A} - \frac23 F_{A}^{\lambda}, \\
\mathcal C_{0b} &=& F_{B} + \frac{ 10 F_{B}^{\lambda}}{3}, \qquad
\mathcal C_{1b} = F_{B} - \frac23 F_{B}^{\lambda}.\label{eq:LaLO2}
\end{eqnarray}
The LO Lagrangian determines the contact interaction potential $V\equiv\Pi_{i=1}^{4}\sqrt{2m_i}\,\tilde V$, with $m_i$ the masses of the involved 4 mesons and $ \tilde V=-\langle\mathcal{L}\rangle$,\footnote{Here $\langle\mathcal{L}\rangle$ represents the element of the Lagrangian sandwiched between the initial and final states of the scattering vertex.} in all strangeness ($\pm 1$, 0 and hidden), isospin and $J^{PC}$ sectors, which will be used below.\footnote{The contact terms for all SU(3) charm-anticharm meson pairs were first obtained in Ref.~\cite{Hidalgo-Duque:2012rqv}.}  Later the potential 
is  used as the kernel of the two-body elastic LSE.

At LO, it is also quite convenient to use the two-component notation for the heavy meson superfield~\cite{Manohar:2000dt,Mehen:2011yh}; see Appendix~\ref{app}.

\section{Extraction of the pole position from the \texorpdfstring{$\bm{\dsds}$}{Ds+Ds-} invariant mass distribution}\label{Sect:fitDsDs}

Although the mass in Eq.~\eqref{eq:lhcb} is above the $\dsds$ threshold, $2m_{D_s} = (3936.70\pm0.14)$~MeV, the peak in the immediate vicinity of the $\dsds$ threshold indicates that it may also be compatible with a below-threshold pole, which can be either a virtual or bound state.

From the LHCb measured spectrum, we can see that no additional structures in $D_s^{\pm}K^\pm$ channels are needed.\footnote{In fact, the higher $D_0^*$ state with a mass about 2.45~GeV claimed in Refs.~\cite{Guo:2006fu,Albaladejo:2016lbb} couples dominantly to the $D_s\bar K$. It was shown in Ref.~\cite{Du:2017zvv} that the existence of such a state and a lower $D_0^*$ one, with a mass around 2.1~GeV, is consistent with the LHCb data on the $D\pi$ angular moments~\cite{LHCb:2016lxy}. One notices that there is some hint of a near-threshold enhancement in the LHCb data of the $D_s^-K^+$ invariant mass distribution~\cite{LHCb:2022NewObservations}. From the Dalitz plot of the $B^-\to D_s^+D_s^-K^+$, it is easy to see that such  $D_0^*$ resonance does not affect the near-threshold $D_s^+D_s^-$ distribution.} Therefore, among the final particles,  we only consider the $D_s^+D_s^-$ re-scattering , and the amplitude for $B^-\to D_s^+D_s^-K^+$ reads
\begin{align}
    T_B(m_{D_s^+D_s^-})={\mathcal P}+{\mathcal P}\,G(m_{D_s^+D_s^-})T(m_{D_s^+D_s^-}) ,
\end{align}
where ${\mathcal P}$ denotes the point-like production source, assumed to be constant and $m_{D_s^+D_s^-}$ is the invariant mass of the $D^+_sD_s^-$ pair.  In addition, $G(W)$ is the two-point loop function with two intermediate mesons, defined by 
%
\begin{align}
    G(W)=i\int \frac{d^4q}{(2\pi)^4}\frac1{(q^2-m_{1}^2+i\epsilon)[(P-q)^2-m_{2}^2+i\epsilon]} ,
    \label{eq:G}
\end{align}
with $m_1,m_2$ the masses of the intermediate particles, which are the $D_s^+D_s^-$ pair in this section, and $P$ their four-momentum [$P^\mu = (W,\vec 0\,)$ in the two-meson c.m. system]. 
Using dimensional regularization (DR), it reads
\begin{align}
    G_{\rm DR}(W)=&\frac{ 1}{16\pi^2}\bigg\{a(\mu)+\log\frac{m_{1}^2}{\mu^2}+\frac{m_{2}^2-m_{1}^2+s}{2s} \log\frac{m_{2}^2}{m_{1}^2} \nonumber\\
&+\frac{k}{W} \Big[ 
\log\left(2k W+s+\Delta\right) + 
\log\left(2k W+s-\Delta\right) \nonumber\\ 
& -  
\log\left(2k W-s+\Delta\right) - 
\log\left(2k W-s-\Delta\right)
\Big]\bigg\},\label{eq:GDR}
\end{align}
where $s=W^2$, $\Delta= m_{1}^2-m_{2}^2$, $k=\lambda^{1/2}(W^2,m_{1}^2,m_{2}^2)/(2W)$ is the corresponding three-momentum with $\lambda(x,y,z)=x^2+y^2+z^2 - 2xy - 2yz - 2xz$  the K\"all\'en triangle function, and $a(\mu)$ is a subtraction constant 
with  $\mu$, chosen to be 1 GeV,   the DR scale. The branch cut of $k$, taken from the threshold to infinity along the positive real $W$ axis, splits the whole complex energy ($W$) plane into two RSs defined as Im$(k)>0$ on the first RS and Im$(k)<0$ on the second RS. Another way to regularize the loop integral is inserting a Gaussian form factor, namely,
\begin{align}
 G_{\Lambda}(W) =&\, \frac{1}{4m_1m_2}
 \int \frac{l^2 dl}{2\pi^2}  
 \frac{e^{-2l^2/\Lambda^2} }{W-l^2/2\mu_{12}-m_1-m_2+i\epsilon},\label{eq:GGF}
\end{align}
with $\mu_{12}=m_1m_2/(m_1+m_2)$ the reduced mass, where the nonrelativistic approximation has been taken to both intermediate particles. The cutoff $\Lambda$ is usually in the range of $0.5\sim 1.0$ GeV. The subtraction constant $a(\mu)$ in DR is determined by matching the values of the loop function $G$ obtained from these two methods at threshold, $W=(m_1+
m_2)$. We will use the DR loop with the so-determined subtraction constant for numerical calculations.

$T(m_{D_s^+D_s^-})$ is the amplitude for the $D^+_sD_s^-$ elastic scattering, and it is given by
\begin{align}
    T(m_{D_s^+D_s^-})=\frac{{V}}{1-{V} G(m_{D_s^+D_s^-})},
\end{align}
with $V=4m_{D_s}^2(\mathcal C_{0a}+\mathcal C_{1a})/2$, 
the potential for the $D^+_sD_s^-$ system~\cite{Hidalgo-Duque:2012rqv}. The invariant mass distribution of $D^+_sD_s^-$ is described by
\begin{align}
    \frac{{\rm d} \Gamma}{{\rm d} m_{D_s^+D_s^-}}=\frac{1}{(2\pi)^3}\frac{k\, p}{4m_B^2}|T_B(m_{D_s^+D_s^-})|^2,
\end{align}
with $k$ the three-momentum of the $K^+$ in the rest frame of the $B^+$ meson, $p$ the three-momentum of the $D_s^+$ in the c.m. frame of the $D_s^+D_s^-$ pair. Averaging the differential decay width for each experimental bin and fitting to  the LHCb data, we obtain the interaction strength of $D_s^+D_s^-$ and, in turn, the pole position of this system.

The fitted distribution with $\Lambda=0.5$ GeV is shown in Fig.~\ref{fig:fit_DsDs}. Actually there are two solutions, one which develops a bound state pole, and the other one with a virtual state pole, but they are almost indistinguishable, which follows from the fact that there are only data points above the $D_s^+D_s^-$ threshold.\footnote{The pole is located on the real axis below $D_s\bar D_s$ threshold in the single-channel case, and it corresponds to a bound or virtual state of $D_s\bar D_s$. If the couplings to the $D\bar D$ or the OZI suppressed $J/\psi\omega$ lower-energy channels are considered, the pole will move to the complex plane becoming a resonance. Abusing the notation, we will continue to call such pole a bound or virtual state, since it would have essentially the same nature as in the single channel case---it is formed by the attraction between the $D_s\bar D_s$ mesons.} The fit with $\Lambda=1.0$ GeV has the same quality and leads to similar pole positions, as seen in Table~\ref{tab:fit_res}. 
\begin{figure}[tb]
    \centering
    \includegraphics[width=0.80\textwidth]{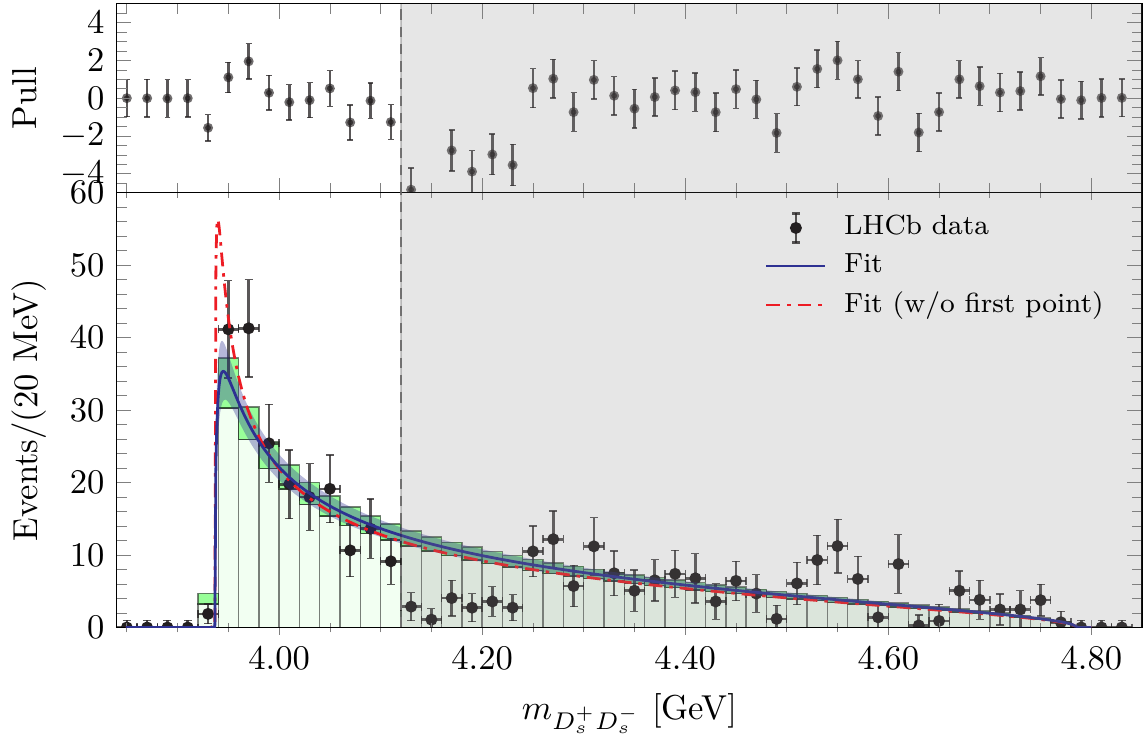}
    \caption{Bottom panel: Fitted line shapes of the $D_s^+D_s^-$ invariant mass distribution in the $B^+\to D_s^+D_s^-K^+$ reaction. Data taken from Ref.~\cite{LHCb:2022NewObservations}. The shaded area on the right, starting at $m_{D_s^+ D_s^-}=4.12\,\text{GeV}$ is the region excluded in all fits. The solid blue line and green histogram stand for the best fit ($\chi^2/{\rm dof}=1.40$) to  the ten non-zero data points included in the non-shaded area in the plot (from 3.92~GeV to 4.12~GeV), which leads to a virtual state pole. The dashed-dotted red line is obtained considering only the 9 points from 3.94~GeV ($\chi^2/{\rm dof}=0.83$). The band around the solid line and the highlighted area in top of each histogram show the uncertainty obtained Monte Carlo propagating the errors from the fit to the data. The curves obtained from fits leading to a bound state pole, as discussed in the text, are almost indistinguishable from those shown here and have an almost identical $\chi^2/\text{dof}$. Top panel: Pull of the data for the fit, defined as the difference between the heights of the experimental and  theoretical histograms divided by the error of the former.
    }
    \label{fig:fit_DsDs}
\end{figure}

\begin{table}[tb]
\renewcommand\arraystretch{1.2}
\begin{ruledtabular}
\caption{Potential and pole positions from different fits to the LHCb data. The poles are located either on the real axis on the first (bound) or the second (virtual) RSs below threshold, and their positions are defined by the distance in MeV from the $D_s^+D_s^-$ threshold, i.e., $M-2 m_{D_s}$. The errors on the pole positions are propagated from the fits.
}\label{tab:fit_res}
\begin{tabular}{l|cc|cc}
    & \multicolumn{2}{c}{$\Lambda=0.5$ GeV}           & \multicolumn{2}{c}{$\Lambda=1.0$ GeV}             \\
           \hline
           & virtual                & bound                  & virtual                & bound                  \\
           \hline
           $V$        & $-300\pm 20$     & $-1300_{-500}^{+300}$  & $-190\pm 10$       & $-360\pm 30$     \\
pole [MeV] & $-8.4^{+2.4}_{-3.2}$ & $-8.4^{+2.3}_{-3.3}$ & $-8.4^{+2.5}_{-3.3}$ & $-8.4^{+2.3}_{-3.3}$
\end{tabular}

\end{ruledtabular}
\end{table}

With the binding or virtual energy given in Table~\ref{tab:fit_res}, we get a mass of the $\dsds$ bound or virtual state of
\begin{align}
    M_{X(3960)} = (3928\pm3)~{\rm MeV}. \label{eq:mass}
\end{align}
The lattice QCD calculation in Ref.~\cite{Prelovsek:2020eiw} predicts the existence of a $\dsds$ bound state with a binding energy in the range of [2.4, 8.2]~MeV,\footnote{The authors of  Ref.~\cite{Prelovsek:2020eiw} obtain two shallow bound states when they consider separately  the $D \bar D$ and $D_s \bar D_s$ single channels, which agree with our results. Then, they find that the masses slightly change by including $D \bar D$--$D_s \bar D_s$ coupled-channel effects, which gives support to perform single-channel analyses. In addition, they also found a third resonance above the $D_s \bar D_s$ threshold, which appears due to the use of an energy-dependent potential, and a $D$-wave $2^{++}$ state. These two latter results are beyond the scope of this work, where we will  
only consider $S$-wave constant interactions. } and the model in Ref.~\cite{Dong:2020hxe}, which assumes that the interaction between a pair of charmed hadrons is dominated by the exchange of light-flavor vector mesons, predicts a $\dsds$ virtual state with a virtual energy in the range of [4.7, 35.5]~MeV. 
Both predictions are consistent with the determinations in Table~\ref{tab:fit_res}.

\section{Determination of the LO Lagrangian $\mathbf{\mathcal L_{4H}}$  }\label{sect:spectrum}
The four independent LECs, $\mathcal C_{0a},\mathcal C_{0b},\mathcal C_{1a}$ and $\mathcal C_{1b}$, in Eqs.~\eqref{eq:LaLO}-\eqref{eq:LaLO2} can be determined from four independent inputs:
\begin{itemize}
    \item The mass of the $X(3872)$ as a bound state of the $D^0\bar D^{*0}$-$D^+D^{*-}$ system. The $D^0\bar D^{*0}$-$D^+D^{*-}$ coupled-channel interaction in the $J^{PC}=1^{++}$ sector reads~\cite{Hidalgo-Duque:2012rqv} 
\begin{align}
        \tilde V=\frac{1}{2}\left(\begin{array}{ll}
        \mathcal C_{0X}+\mathcal C_{1X} & \mathcal C_{0X}-\mathcal C_{1X} \\
        \mathcal C_{0X}-\mathcal C_{1X} & \mathcal C_{0X}+\mathcal C_{1X}
        \end{array}\right), \label{eq:VX}
\end{align}
with $\mathcal C_{0X}=\mathcal C_{0 a}+\mathcal C_{0 b}$, $\mathcal C_{1X}=\mathcal C_{1 a}+\mathcal C_{1 b}$.  We take the mass of $X(3872)$ from a recent LHCb analysis using a Flatt\`e parametrization~\cite{LHCb:2020xds}, 
\begin{align}
M_X & =3871.69_{-0.04-0.13}^{+0.00+0.05} \ \mathrm{MeV}, 
\end{align}
which fixes the binding energy relative to the threshold of $D^0\bar D^{*0}$ to be $M_X-m_{D^0} - m_{D^{*0}} = [-150,0]$ keV.\footnote{With the masses $m_{D^0}=(1864.84\pm0.05)$~MeV and  $m_{D^{*0}}=(2006.85\pm0.05)$~MeV~\cite{Workman:2022ynf}, the $D^0D^{*0}$ threshold is at $(3871.69\pm0.07)$~MeV. Here we consider  that the $X(3872)$ is bound state below the threshold and the lower bound is estimated by a quadratic sum of these errors to be 150 keV.}  We obtain a relation among these four LECs by reproducing the binding energy of the $X(3872)$ relative to the $D^0\bar D^{*0}$ threshold. In this work, the central value of $M_X$ is taken at 3871.69 MeV and to estimate the inherited  uncertainties affecting  the LECs,  we perform a Monte Carlo (MC) sampling according to a uniform distribution between 3871.54 and 3871.69 MeV.
\item The ratio of isospin breaking decays of the $X(3872)$  given by~\cite{Gamermann:2009uq,Hidalgo-Duque:2012rqv,Albaladejo:2015dsa}
\begin{align}\label{eq:ratio}
        R_{X(3872)}=\frac{\hat{\Psi}_{\mathrm{n}}-\hat{\Psi}_{\mathrm{c}}}{\hat{\Psi}_{\mathrm{n}}+\hat{\Psi}_{\mathrm{c}}},
    \end{align}
    where $\hat{\Psi}_{\mathrm{n,c}}$ give the average of the neutral and charged wave
function components in the vicinity of the origin, This ratio is 
related to $\mathcal C_{0X}$ and $\mathcal C_{1X}$ \cite{Gamermann:2009uq}
    \begin{align}\label{eq:ncratio}
        \frac{\hat{\Psi}_{\mathrm{n}}}{\hat{\Psi}_{\mathrm{c}}}=\frac{1-(2m_{D^+}m_{D^{*-}})\,G_{2}\left(\mathcal C_{0 X}+\mathcal C_{1 X}\right) }{(2m_{D^+}m_{D^{*-}})\,G_{2}\left(\mathcal C_{0 X}-\mathcal C_{1 X}\right)}=\frac{(2m_{D^0}m_{D^{*0}})\,G_{1}\left(\mathcal C_{0 X}-\mathcal C_{1 X}\right)}{1-(2m_{D^0}m_{D^{*0}})\,G_{1}\left(\mathcal C_{0 X}+\mathcal C_{1 X}\right)},
    \end{align}
    with $G_1$ and $G_2$ the loop functions of $D^0\bar D^{*0}$ and $D^+ D^{*-}$, respectively, evaluated at the pole position of the $X(3872)$. The experimental value of the ratio is $0.29\pm 0.04$~\cite{LHCb:2022bly} and a MC sampling according to the Gaussian distribution with a mean value of $0.29$ and a standard deviation of $0.04$ is performed to estimate the uncertainties on the derived  LECs.
    
    \item The pole position of the $Z_{c}(3900)$. The contact interaction strength for this system, the isovector $D\bar D^*+\bar D D^*$, is { $ \tilde V=\mathcal C_{1a}-\mathcal C_{1b}$}.  We use for the pole position, $3813^{+28}_{-21}$~MeV, from the virtual state pole scenario discussed in Ref.~\cite{Du:2022jjv}, in the context of a combined fit to the BESIII data on the $Z_c(3900)$ and $Z_{cs}(3985)$.\footnote{In Ref.~\cite{Du:2022jjv}, there are two types of fits, leading to a resonance or a virtual state pole for the $Z_c(3900)$. The virtual state scenario is the one consistent with the LO NRFET approach followed here (see Eq.~\eqref{eq:LaLO}).} A MC sampling of the pole position according to a Gaussian distribution with mean value of $3813$ MeV and standard deviation of $28$ MeV is performed to estimate the uncertainties of the LECs.
    
    \item The $D_s^+D_s^-$ interaction strength, i.e.  $\tilde V=(\mathcal C_{0a}+\mathcal C_{1a})/2$ , obtained in Sect.~\ref{Sect:fitDsDs}. We directly use the potentials listed in Table~\ref{tab:fit_res} with Gaussian distributions to estimate the uncertainties of the derived LECs. 
\end{itemize}

For each two-body channel, the subtraction constant in the DR loop function is fixed by  matching the values of  Eqs.~\eqref{eq:GDR} and \eqref{eq:GGF} at the corresponding threshold.

From the $X(3872)$ inputs, we obtain for $\Lambda=0.5\, (1.0)$ GeV,
\begin{align}
    \mathcal{C}_{0X}&=-1.96_{-0.08}^{+0.03}\ (-0.73_{ -0.02}^{+0.01})\ \rm{fm}^2,\\ \mathcal{C}_{1X}&=1.14\pm0.49 \ (-0.29_{-0.08}^{+0.06})\ \rm{fm}^2. \label{eqc0x}
\end{align}
Note that there are sizeable differences between the above values and those obtained in Ref.~\cite{Albaladejo:2015dsa}, which is traced to the updated experimental meson masses and the isospin-breaking  ratio $R_{X(3872)}$, and the use in this work of the DR scheme in coupled channels.\footnote{From Eqs.~\eqref{eq:ratio} and \eqref{eq:ncratio}, one finds
\begin{equation}
 \mathcal C_{1X}=  -\mathcal C_{0X} R^{-1}_{X(3872)}+\frac{1+R_{X(3872)}^{-1}}{4m_{D^0}m_{D^{*0}}\,G_1}
\end{equation}
The difference between the $\mathcal{C}_{0X}$ values determined using either the Gaussian or the DR schemes is around $0.3$ fm$^2$ for $\Lambda=0.5$ GeV, which induces a large change of the order of $-1$~fm$^2$ for $\mathcal{C}_{1X}$, since it is enhanced by the $R_{X(3872)}^{-1}$ factor.} 
From the last two inputs, which are the BESIII $Z_{c}(3900)$ data and the newly reported LHCb $X(3960$), we have
\begin{align}
    \mathcal C_{1a}-\mathcal C_{1b}&=-0.43_{-0.09}^{+0.06}\ \left(-0.32_{-0.05}^{+0.03}\,\right)\ \rm{fm}^2,\\
    (\mathcal C_{0a}+\mathcal C_{1a})/2&=\left\{\begin{array}{ll}
       -0.75\pm0.05\ (-0.48\pm 0.03)\ \rm{fm}^2  & \text{in S-I}, \\
         -3.3_{-1.3}^{+1.2}\ (-0.9\pm 0.1)\ \rm{fm}^2  & \text{in S-II},
    \end{array}\right.
\end{align}
where we have considered two scenarios I (S-I) and II (S-II), depending on whether the rescattering of the $\dsds$ pair gives rise to  a virtual or bound state pole, respectively. We finally obtain the four LECs,
\begin{align}
    \mathcal C_{0a}&=\left\{\begin{array}{ll}
       -1.85^{+0.28}_{-0.26}\ \left(-0.65_{-0.06}^{+0.08}\,\right)\ \rm{fm}^2  &\text{in S-I},   \\
        -6.9\pm2.5\ \left(-1.50_{-0.15}^{+0.17}\,\right)\ \rm{fm}^2 &  \text{in  S-II}, 
    \end{array}\right.\\
    \mathcal C_{0b}&=\left\{\begin{array}{ll}
     -0.11^{+0.22}_{-0.29}\ (-0.08
_{-0.08}^{+0.05})\ \rm{fm}^2 & \text{in S-I},  \\
        4.9\pm2.5\ \left(0.77_{-0.17}^{+0.14}\,\right)\ \rm{fm}^2&  \text{in S-II},
    \end{array}\right.\\
    \mathcal C_{1a}&=0.36^{+0.25}_{-0.26}\ \left(-0.31_{-0.05}^{+0.03}\,\right)\ \rm{fm}^2,\\
    \mathcal C_{1b}&=0.78^{+0.26}_{-0.24}\ \left(0.01_{-0.05}^{+0.04}\,\right)\ \rm{fm}^2.
\end{align}

\begin{table}[tbp]
  \caption{Predicted pole positions of the SU(2) isoscalar HQSS partners of the $X(3872)$ resonance for two different values
of the Gaussian cutoff. The energies, with $|E|$ the distance between the pole mass and the corresponding threshold, are given in MeV and, a negative (positive) number means a bound (virtual) state located on the first (second) RS {\em below} threshold. For each channel, the first and second rows of values for $E$ stand for the predictions within the S-I and S-II schemes, respectively.  The first set of errors on $E$ is estimated by the MC sampling of the LECs, while the second one (in brackets) is { the {\em full} error} obtained by including an additional 30\% uncertainty into the LECs to account for HQSS and SU(3) flavor symmetry breaking corrections. Thus, the latter set provides a conservative total error on the pole positions. The thresholds marked by an asterisk (*) are the isospin averaged ones. 
} \label{tab:I=0}
\renewcommand\arraystretch{1.5}
    \centering
    \begin{ruledtabular}
    \begin{tabular}{lccccc}
$J^{P C}$ & ${H\bar H}$ &  $\tilde V$ & $E(\Lambda=0.5\ \mathrm{GeV})$ & $E(\Lambda=1\ \mathrm{GeV})$ &  Threshold [MeV] \\
\hline \multirow{2}{*}{$0^{++}$} & \multirow{2}{*}{$D \bar{D}$}  & \multirow{2}{*}{$\mathcal C_{0 a}$} & $-1.9\pm 1.3(_{-3.0}^{+2.0})$ & $0.0_{-0.4}^{+1.7}(_{-4.3}^{+20})$ &\multirow{2}{*}{$3734.5^*$} \\
&&&$-16_{-2}^{+4}(_{-3}^{+6})$&$-32_{-5}^{+6}(_{-14}^{+20})$&\\
\hline $1^{++}$ & $D^{*} \bar{D}$ & Eq.(\ref{eq:VX}) & Input & Input & ${3871.69} / 3879.91$ \\
\hline
\multirow{2}{*}{$1^{+-}$} & \multirow{2}{*}{$D^{*} \bar{D}$}  & \multirow{2}{*}{$\mathcal C_{0 a}-\mathcal C_{0 b}$} &$-1.6_{-2.4}^{+1.7}(_{-3.4}^{+3.9})$ &$ 1.1_{-1.4}^{+20}(_{-3.9}^{+66})$  & \multirow{2}{*}{$3875.8^{*}$} \\
&&&$-19_{-1}^{+3}(_{-2}^{+4})$&$-52_{-5}^{+7}(_{-10}^{+15})$&\\
\hline 
\multirow{2}{*}{$0^{++}$} & \multirow{2}{*}{$D^{*} \bar{D}^{*}$}  & \multirow{2}{*}{$\mathcal C_{0 a}-2 \mathcal C_{0 b}$} &$-1.3_{-3.4}^{+6.6}(_{-4.1}^{+11})$ &$ 5.0_{-\phantom{1}5.0}^{+100}(_{-\phantom{1}7.0}^{+190})$ & \multirow{2}{*}{$4017.1^{*}$} \\
&&&$-19_{-1}^{+2}(_{-1}^{+3})$&$-63_{-4}^{+7}(_{-\phantom{1}7}^{+12})$&\\
\hline
\multirow{2}{*}{$1^{+-}$} & \multirow{2}{*}{$D^{*} \bar{D}^{*}$}  &\multirow{2}{*}{ $\mathcal C_{0 a}-\mathcal C_{0 b}$} &$-1.8_{-2.4}^{+1.9}(_{-3.4}^{+3.5})$ &$ 0.5_{-1.2}^{+16}(_{-4.1}^{+58})$ & \multirow{2}{*}{ $4017.1^{*}$} \\
&&&$-18_{-1}^{+3}(_{-1}^{+3})$&$-52_{-5}^{+7}(_{-\phantom{0}9}^{+14})$&\\
\hline
\multirow{2}{*}{$2^{++}$} & \multirow{2}{*}{$D^{*} \bar{D}^{*}$} &  \multirow{2}{*}{$\mathcal C_{0 a}+\mathcal C_{0 b}$} & $-3.0_{-0.4}^{+0.1}(_{-2.8}^{+2.5})$ &$-2.3_{-0.5}^{+0.1}(_{-7.9}^{+3.4})$ & \multirow{2}{*}{$4017.1^{*}$} \\
&&&$-3.0_{-0.4}^{+0.1}(_{-8.6}^{+14})$&$-2.3_{-0.5}^{+0.1}(_{-20}^{+50})$&\\
\end{tabular}
\end{ruledtabular}
\end{table}

\begin{table}[tbhp]
  \caption{Same as Table~\ref{tab:I=0}, but for  the hidden-strangeness isoscalar HQSS and light-flavor partners of the $X(3872)$ resonance.}
    \label{tab:hiddens}
\renewcommand\arraystretch{1.5}
    \centering
    \begin{ruledtabular}
    \begin{tabular}{l c c c c c}
$J^{P C}$ & $H\bar H$  & $\tilde V$ & $E(\Lambda=0.5\ \mathrm{GeV})$ & $E(\Lambda=1\ \mathrm{GeV})$ & Threshold [MeV] \\
\hline $0^{++}$ & $D_{s} \bar{D}_{s}$ &  $\frac{1}{2}\left(\mathcal C_{0 a}+\mathcal C_{1 a}\right)$ & input & input & $3936.7$ \\
\hline \multirow{2}{*}{$1^{++}$} & \multirow{2}{*}{$D_{s}^{*} \bar{D}_{s}$} &  \multirow{2}{*}{$\frac{1}{2}\left(\mathcal C_{0 a}+\mathcal C_{1 a}+\mathcal C_{0 b}+\mathcal C_{1 b}\right)$} & $58_{-\phantom{0}46}^{+170}(_{-\phantom{0}54}^{+320})$ & $2.3_{-1.8}^{+2.1}(_{-2.5}^{+11})$ & \multirow{2}{*}{$4080.5$}\\
&&&$58_{-\phantom{0}46}^{+170}(_{-\phantom{0}60}^{+470})$&$2.3_{-1.8}^{+2.2}(_{-6.5}^{+78})$&\\
\hline
\multirow{2}{*}{$1^{+-}$} & \multirow{2}{*}{$D_{s}^{*} \bar{D}_{s}$} & \multirow{2}{*}{$\frac{1}{2}\left(\mathcal C_{0 a}+\mathcal C_{1 a}-\mathcal C_{0 b}-\mathcal C_{1 b}\right)$} & $0.2_{-0.4}^{+4.0}(_{-0.8}^{+11})$ & $9.3_{-5.7}^{+17}(_{-8.7}^{+47})$ &\multirow{2}{*}{ $4080.5$} \\
&&&$-14_{-2}^{+4}(_{-2}^{+5})$&$-26_{-5}^{+6}(_{-10}^{+12})$&\\
\hline \multirow{2}{*}{$0^{++}$} & \multirow{2}{*}{$D_{s}^{*} \bar{D}_{s}^{*}$} & \multirow{2}{*}{ $\frac{1}{2}\left(\mathcal C_{0 a}+\mathcal C_{1 a}-2 \mathcal C_{0 b}-2 \mathcal C_{1 b}\right)$} & $-0.6_{-2.3}^{+2.2}(_{-2.8}^{+4.6})$ & $12_{-10}^{+51}(_{-12}^{+96})$ & \multirow{2}{*}{ $4224.4$} \\
&&&$-16_{-1.5}^{+3.3}(_{-1.6}^{+3.8})$&$-38_{-5}^{+7}(_{-\phantom{1}9}^{+13})$&\\
\hline
\multirow{2}{*}{$1^{+-}$} & \multirow{2}{*}{$D_{s}^{*} \bar{D}_{s}^{*}$} & \multirow{2}{*}{ $\frac{1}{2}\left(\mathcal C_{0 a}+\mathcal C_{1 a}-\mathcal C_{0 b}-\mathcal C_{1 b}\right)$} & $0.1_{-0.4}^{+3.2}(_{-0.9}^{+9.5})$ & $6.8_{-4.6}^{+14}(_{-6.7}^{+41})$ &\multirow{2}{*}{$4224.4$} \\
&&&$-14_{-2}^{+4}(_{-2}^{+5})$&$-26_{-5}^{+6}(_{-\phantom{1}9}^{+12})$&\\
\hline
\multirow{2}{*}{$2^{++}$} & \multirow{2}{*}{$D_{s}^{*} \bar{D}_{s}^{*}$} & \multirow{2}{*}{$\frac{1}{2}\left(\mathcal C_{0 a}+\mathcal C_{1 a}+\mathcal C_{0 b}+\mathcal C_{1 b}\right)$} & $51_{-\phantom{1}41}^{+160}(_{-\phantom{1}48}^{+310})$ & $1.3_{-1.2}^{+1.6}(_{-1.9}^{+11})$& \multirow{2}{*}{$4224.4$} \\
&&&$51_{-\phantom{1}41}^{+160}(_{-\phantom{1}53}^{+470})$&$1.3_{-1.2}^{+1.6}(_{-6.5}^{+69})$&\\
\end{tabular}
\end{ruledtabular}
\end{table}

\begin{table}[tbhp]
  \caption{Same as Table~\ref{tab:I=0}, but for the isospinor ($I=1/2$) HQSS and light-flavor partners of the $X(3872)$ resonance. In these sectors, the interactions are independent of $\mathcal C_{0a}$ and $\mathcal C_{0b}$, and hence the spectrum is the same for both S-I and S-II schemes. In addition, the numbers marked with $^\dagger$ stand for the averaged value of the two involved thresholds. We use the ``$-$" symbol on those cases, where we find neither a bound nor a virtual pole close to threshold.
  }
    \label{tab:I=1/2}
\renewcommand\arraystretch{1.5}
    \centering
    \begin{ruledtabular}
    \begin{tabular}{l  c c c c c}
$J^{P}$ & $H\bar H$ & $\tilde V$ & $E(\Lambda=0.5\ \mathrm{GeV})$ & $E(\Lambda=1\ \mathrm{GeV})$  & Threshold [MeV] \\
\hline $0^{+}$ & $D_{s} \bar{D}$ &$\mathcal C_{1 a}$ & $-$ &$76_{-32}^{+35}(_{-\phantom{1}55}^{+130})$ & $3835.6^*$ \\
\hline $1^{+}$ & $D_{s} \bar{D}^{*}+ D_{s}^{*} \bar{D}$  & $\mathcal C_{1 a}- \mathcal C_{1 b}$ & $57^{+25}_{-26}(_{-\phantom{1}48}^{+240})$ & $ 56\pm26(_{-\phantom{1}42}^{+110})$ & $3978.2^\dagger$ \\
\hline $1^{+}$ & $D_{s} \bar{D}^{*}- D_{s}^{*} \bar{D}$ & $\mathcal C_{1 a}+ \mathcal C_{1 b}$ & $-$ &  $76_{-47}^{+79}(_{-\phantom{1}60}^{+160})$ & $3878.2^\dagger$ \\
\hline $0^{+}$ & $D_{s}^{*} \bar{D}^{*}$ & $\mathcal C_{1 a}-2 \mathcal C_{1 b}$ & $0.0_{-0.9}^{+1.2}(_{-2.2}^{+10})$& $40_{-28}^{+52}(_{-\phantom{1}34}^{+120})$& $4120.8^{*}$ \\
\hline
$1^{+}$ & $D_{s}^{*} \bar{D}^{*}$& $\mathcal C_{1 a}-\mathcal C_{1 b}$ & $50_{-23}^{+24}(_{-\phantom{1}43}^{+220})$&$47\pm23(_{-\phantom{1}37}^{+100})$& $4120.8^{*}$ \\
\hline
$2^{+}$ & $D_{s}^{*} \bar{D}^{*}$ & $\mathcal C_{1 a}+\mathcal C_{1 b}$ &$-$ & $65_{-42}^{+72}(_{-\phantom{1}53}^{+150})$&  $4120.8^{*}$\\
\end{tabular}
\end{ruledtabular}
\end{table}

\begin{table}[tbhp]
 \caption{Same as Table~\ref{tab:I=1/2}, but for the  isovector HQSS partners of the $X(3872)$ resonance.  Note that the isovector $1^{++}$ $D^*\bar D$ channel  is directly related to the isospin breaking  in the $X(3872)$  decays.
 }
    \label{tab:isovector}
\renewcommand\arraystretch{1.5}
    \centering
    \begin{ruledtabular}
    \begin{tabular}{l  c c c c c}
$J^{P C}$ &  $H\bar H$ & $\tilde V$ & $E(\Lambda=0.5\ \mathrm{GeV})$ & $E(\Lambda=1\ \mathrm{GeV})$ & Threshold [MeV] \\
\hline $0^{++}$ & $D \bar{D}$  & $\mathcal C_{1 a}$ & $-$ & $84_{-35}^{+37}(_{-\phantom{1}60}^{+130})$  & $3734.5^*$ \\
\hline $1^{++}$ & $D^{*} \bar{D}$ & Eq.(\ref{eq:VX}) & $-$ &$-$  & $3871.68 / 3879.91$\\
\hline
$1^{+-}$ & $D^{*} \bar{D}$  & $\mathcal C_{1 a}-\mathcal C_{1 b}$ & Input & Input &  $3875.8^{*}$ \\
\hline $0^{++}$ & $D^{*} \bar{D}^{*}$  & $\mathcal C_{1 a}-2 \mathcal C_{1 b}$ & $0.0_{-0.8}^{+1.6}(_{-2.1}^{+12})$ &  $45_{-30}^{+56}(_{-\phantom{1}37}^{+130})$ & $4017.1^{*}$\\
\hline
$1^{+-}$ & $D^{*} \bar{D}^{*}$  & $\mathcal C_{1 a}-\mathcal C_{1 b}$ & $55\pm 25(_{-\phantom{1}46}^{+230})$ &$53\pm 25(_{-\phantom{1}41}^{+110})$   & $4017.1^{*}$ \\
\hline
$2^{++}$ & $D^{*} \bar{D}^{*}$  & $\mathcal C_{1 a}+\mathcal C_{1 b}$ & $-$ & $73_{-46}^{+76}(_{-\phantom{1}58}^{+160})$ & $4017.1^{*}$ \\
\end{tabular}
\end{ruledtabular}
\end{table}
\vspace{0.25cm}

\begin{figure}
    \centering
    \includegraphics[width=\linewidth]{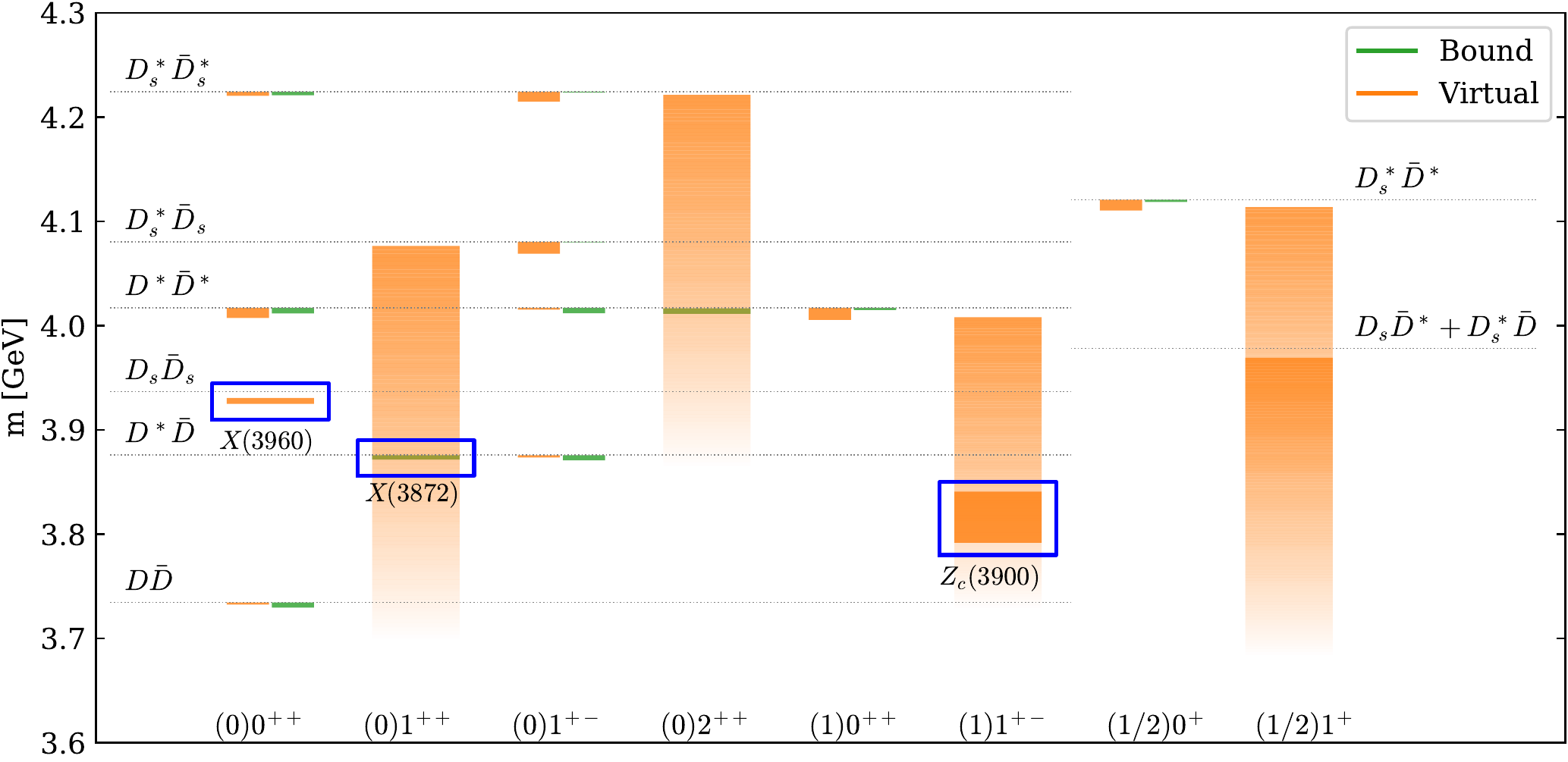}\\
    \includegraphics[width=\linewidth]{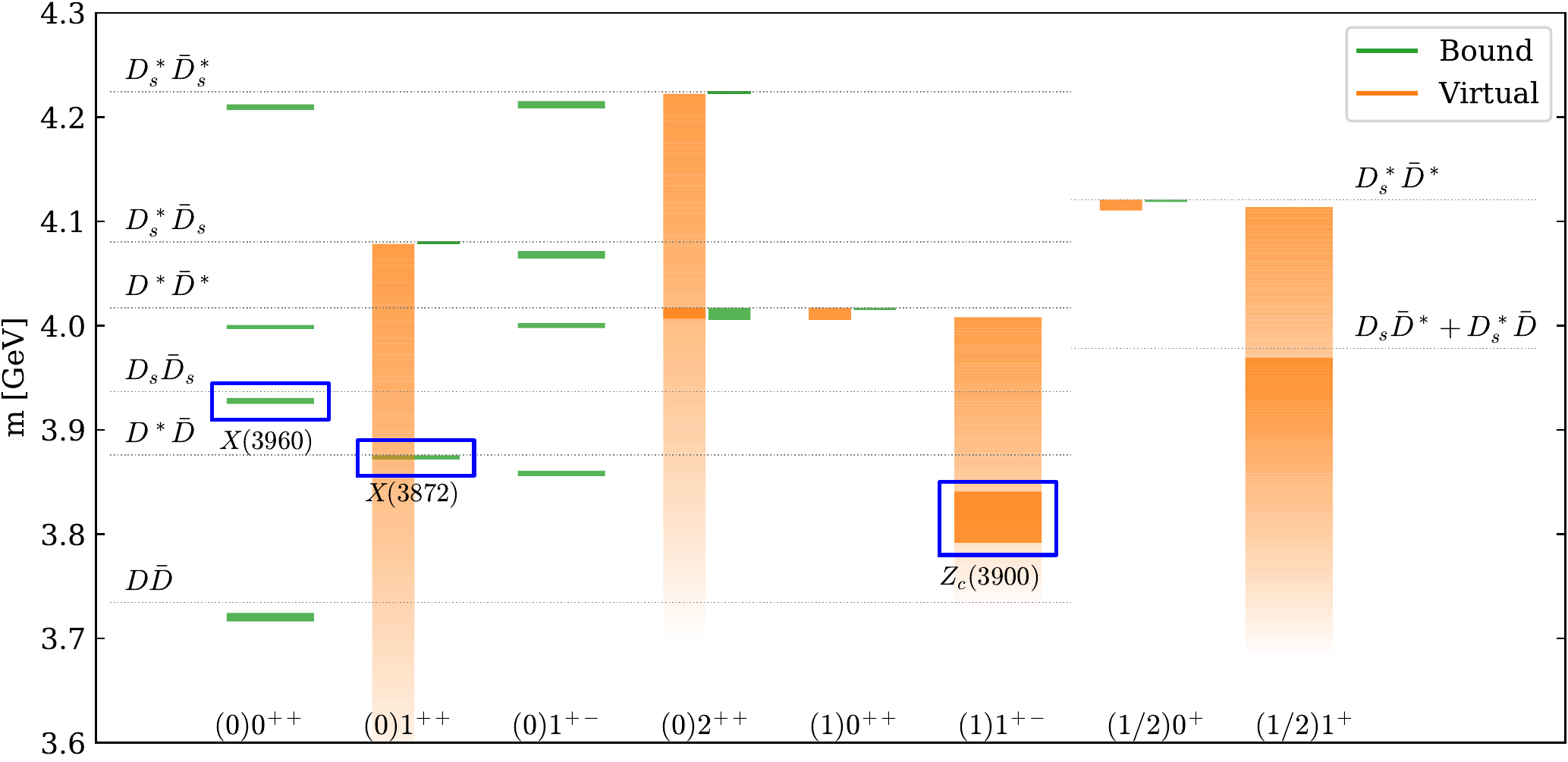}
    \caption{Complete $H \bar{H}$ molecular spectrum   obtained for both the S-I (top) and S-II (bottom) schemes and with a cutoff $\Lambda=0.5$ GeV. Quantum numbers $(I)J^{PC}$ are specified at the bottom of the plots. The filled rectangle bands, green or orange  for bound or virtual states, respectively, cover the range of the pole positions given in Tables~\ref{tab:I=0}, \ref{tab:hiddens}, \ref{tab:I=1/2} and \ref{tab:isovector} (second sets of errors). Thresholds are marked by dotted horizontal lines. The band closest to, but below, the threshold would correspond to a $H\bar H$ hadronic molecule, and with quantum numbers indicated at the bottom. In the cases where the range of the pole position lies in both the bound and virtual regions, we show the two filled shorter bands. The input channels are signaled by a blue rectangle. The band width for the $X(3872)$ is multiplied by a factor of 10 for a better illustration. Note that in some cases the error for the pole position is so large that exceeds the validity region of the NREFT treatment, and hence the bands fade away.}
    \label{fig:spec}
\end{figure}

\section{Summary and discussions}

Using the LECs determined in the previous section, we obtain the complete spectrum of the $S$-wave $H\bar H$ hadronic molecules, which is collected in Tables~\ref{tab:I=0}, \ref{tab:hiddens}, \ref{tab:I=1/2} and \ref{tab:isovector} and shown in Fig.~\ref{fig:spec}. Some further remarks on the  estimation of errors in these tables are given in  Appendix \ref{app:errors}.
We see that several HQSS/SU(3) siblings of the $X(3872)$, $Z_c(3900)$ and $X(3960)$ resonances are predicted. Some of them might not be real QCD bound states, but rather virtual state poles, which will nevertheless produce some nontrivial structures in the experimental distributions of $H\bar H$ or other final states that can couple to $H\bar H$ (for a discussion on some general features of the virtual state line shapes, see Ref.~\cite{Dong:2020hxe}).

The SU(3) content of the found spectrum is determined from the flavour decomposition $[\bar 3] \otimes [3] = [1] \oplus [8]$. Let us pay, for instance, attention to the pseudoescalar-vector $J^{PC}=1^{+-}$  sector.  The octet interaction can be read off from that in the   $I=1$ or $I=1/2$ isospin channels~\cite{Hidalgo-Duque:2012rqv, Yang:2020nrt,Du:2022jjv} and it is determined by the linear combination $[\mathcal C_{1a}-\mathcal C_{1b}]$ (see Tables~\ref{tab:I=1/2} and \ref{tab:isovector}). Hence, the obtained molecular-like exotic states $Z_c(3900)$ and $Z_{cs}(3985)$ belong to the same octet~\cite{Yang:2020nrt,Du:2022jjv}, as depicted in Fig.~\ref{fig:octet}. For $I=0$, one has the LECs   $(\mathcal C_{0a}-\mathcal C_{0b})$  and $\frac{1}{2} \left( \mathcal C_{0a} - \mathcal C_{0b} + \mathcal C_{1a}-\mathcal C_{1b} \right)$ for the $D^\ast \bar{D}$ and $D_s^* \bar{D}_s$ (hidden strangeness) pairs, respectively. In the SU(3) limit, it would be necessary to account for the coupled-channel dynamics, and  the $I=0$ potential would diagonalize into octet and  singlet components. The eigenvalue of the octet piece  would be $(\mathcal C_{1a} - \mathcal C_{1b})$, and it  would give rise to the $I=0$ member of the octet in Fig.~\ref{fig:octet}. Hence due the single-channel scheme used here, appropriate because of the sizable $D^{(*)}_s-D^{(*)}$ mass splitting (see discussion below), neither the $D^\ast \bar{D}$ nor the $D_s^* \bar{D}_s$  $(0)1^{+-}$ molecular-like states reported in Tables~\ref{tab:I=0} and \ref{tab:hiddens} can be directly identified with the SU(3)  octet and singlet isoscalar states missing in Fig.~\ref{fig:octet}. The former ones will be linear combinations of the SU(3) flavor eigenstates, similarly to that, for instance, the physical $\omega$ and $\phi$ mesons are obtained from the mixing of the isoscalar SU(3) singlet and octet states. 

One would find a second $J^{PC}=1^{+-}$ nonet (octet+singlet) of HQSS siblings of the latter one in the vector-vector  sector, where the charged $Z^*_c(4020)$ and $Z^*_{cs}$ would be located. The same SU(3)/HQSS pattern would be repeated for the $J^{PC}=1^{++}$  pseudoscalar-vector and $J^{PC}=2^{++}$ vector-vector systems,  and finally there would be   two (pseudoscalar-pseudoscalar and  vector-vector) independent $J^{PC}=0^{++}$ nonets. Nevertheless, in some cases the interaction might not be strong enough to even generate a virtual state,  close enough to the relevant threshold, that can produce non-negligible signatures in observable distributions.

\begin{figure}
    \centering
    \includegraphics[height=5cm,keepaspectratio]{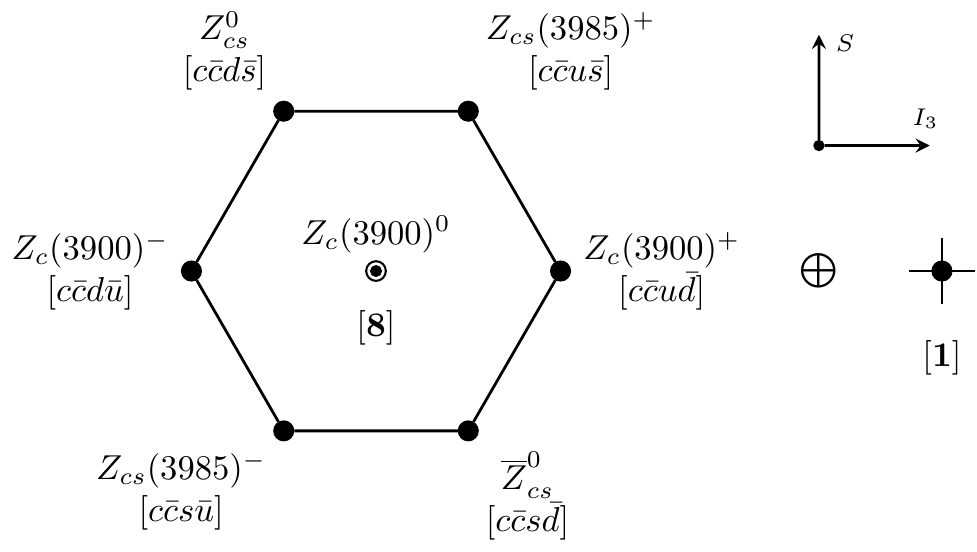}
    \caption{SU(3) content of the pseudoescalar-vector $J^{PC}=1^{+-}$  molecular sector, deduced from the $[\bar 3] \otimes [3] = [1] \oplus [8]$ reduction. See text for details.}
    \label{fig:octet}
\end{figure}

To summarize, we have shown that the freshly measured $\dsds$ invariant mass distribution in the $B^+\to\dsds K^+$ reaction, reported by the LHCb Collaboration~\cite{LHCb:2022NewObservations}, can be well described by a $\dsds$ bound or virtual state below threshold. The pole is located at $(3928\pm3)$~MeV on the first or second RS for a bound or a virtual state, respectively.
Thus, the observed peak in the data above threshold can be just a consequence of the phase-space suppression.
In this and other similar situations, the line shape in the near-threshold region should be analyzed using a parametrization which can be analytically continued to the below-threshold region (a nice example in this context is provided by the LHCb analysis of the $X(3872)$ line shape using the Flatt\`e parametrization~\cite{LHCb:2020xds}). The pole position obtained from the analytically-extended  amplitude in such analyses is always more meaningful than the BW parameters.

This possible new $\dsds$ molecule [$X(3960)$], together with inputs from the mass and isospin breaking decays of the $X(3872)$, and the virtual state pole position of the $Z_c(3900)$ determined in Ref.~\cite{Du:2022jjv}, have allowed us to fully fix the four LECs that appear in  the $4H$ NREFT Lagrangian at LO ($\mathcal L_{4H}$). Next, we have used this contact-range interaction to predict the  complete spectrum of  hadronic molecules formed by an $S$-wave pair of ground state charmed and anticharmed mesons. It is expected that coupling channels, with the same quantum numbers and thresholds more than 100~MeV away from each other, can bring in corrections to the numerical results, but the general pattern should hold; see the analyses in Ref.~\cite{Nieves:2012tt} for $H\bar H$ systems in the SU(2) flavor case and Ref.~\cite{Du:2021fmf} for the hidden-charm pentaquarks. A similar remark applies to the effects of pion exchanges, which are also not considered in this work.

We have regarded near-threshold virtual states as molecular states despite that their wave functions are not normalizable.
It is worthwhile to notice that the wave function of a resonance, whose poles are located on the unphysical Riemann sheet of the complex energy plane just like a virtual state pole, is not normalizable either. Although both virtual states and resonances are not asymptotic states, they are able to produce observable signatures, such as peaks in  invariant mass distributions, as long as the poles are not far from the physical region. In particular, it is important to stress that the pole position of a virtual state, as that of a resonance or a bound state, is the same in every reaction with the same quantum numbers and initial or final state.
Thus, a virtual state may be regarded as a particle just like what is normally done for a resonance on the same footing.

Notice that we have ignored  higher order interactions and coupled-channel dynamics. In principle, coupled channels may have some effects on the spectrum. However, in this work we have mainly focused on  poles near thresholds, for which only considering single-channel dynamics should be a good approximation, since the energy-distance to the corresponding threshold is much smaller than the separation to the closest energy channel.\footnote{The typical momentum
scale of the coupled channel is $\sqrt{2\mu m_\pi}\sim$  500 MeV, with $\mu$ the heavy-light meson pair reduced mass and $(m_{D^*}-m_D) \sim m_\pi$, which turns out to be a hard scale of the order of the ultraviolet cutoff, and much larger than the binding momentum of a shallow $S$-wave bound or virtual state. Note that the $D_s-D$ mass splitting, though smaller than  $(m_{D^*}-m_D)$, is still of around 100 MeV. On the other hand, we should point out that the lowest-order counterterm structure stemming from the NREFT cannot absorb the kind of divergences associated with the coupled channel calculations and one would need to introduce higher-order new counterterms to renormalize the coupled channel dynamics. These problems were addressed in Ref.~\cite{Nieves:2012tt}, and shown there that in any case the effects should be safely covered by uncertainties coming from the $1/m_Q$ corrections. See also Ref.~\cite{Du:2021fmf} for discussions in the case of the pentaquark $P_c$. } In this case, the poles are always located at real axis below the threshold. If we had taken into account lower-energy channels, the near threshold poles would have only moved to the complex plane, which rarely affects the mass of the state. 

Molecular states can also in principle couple to near-threshold $q \bar q$ components if their quantum numbers are the same. The recent analysis in Ref.~\cite{Hanhart:2022qxq} suggests that it is plausible to treat hadronic molecules isolated from $q\bar q$ states of the same quantum numbers. Nevertheless, the interactions driven by $q \bar q$ compact states located some few tens of MeV below/above thresholds can be incorporated into the EFT by appropriate LECs, when only narrow energy windows around some relevant two-meson thresholds are being considered. This is because  then the energy-dependence of the interaction can be safely neglected. In this context, the study carried out in Ref.~\cite{Cincioglu:2016fkm} for the $X(3872)$ and its $2^{++}$ HQSS partner is very illustrative.

\section*{Acknowledgements}
We are grateful to Bing-Song Zou for helpful discussions as well as his careful reading of this manuscript. This research has been supported  by the Spanish Ministerio de Ciencia e Innovaci\'on (MICINN)
and the European Regional Development Fund (ERDF) under Contract PID2020-112777GB-I00; 
by the EU STRONG-2020 Project under the Program H2020-INFRAIA-2018-1 with 
Grant Agreement No. 824093; by  Generalitat Valenciana under Contract PROMETEO/2020/023; by the Chinese Academy of Sciences under Grant No.~XDB34030000; by the National Natural Science Foundation of China (NSFC) under Grants No.~12125507, No. 11835015, No.~12047503, and No.~11961141012; and by the NSFC and the Deutsche Forschungsgemeinschaft (DFG) through the funds provided to the Sino-German Collaborative Research Center TRR110 “Symmetries and the Emergence of Structure in QCD” (NSFC Grant No.~12070131001, DFG Project-ID~196253076). M.~A. is supported by Generalitat Valenciana under Grant No. CIDEGENT/2020/002.

\begin{appendix}
\section{Lagrangian in the two-component notation}
\label{app}

Taking $v^\mu = (1, \vec 0)$, we have 
\begin{align}
    H_a^{(Q)} = \left(\begin{array}{cc}
0 & -\vec{P}_{a}^{*(Q)} \cdot \vec{\sigma}-P_{a}^{(Q)} \\
0 & 0
\end{array}\right), \qquad
H^{(\bar Q)a} = \left(\begin{array}{cc}
0 & -\vec{P}^{*(\bar Q)a} \cdot \vec{\sigma}-P^{(\bar Q)a} \\
0 & 0
\end{array}\right),
\end{align}
where $\vec \sigma$ denotes the Pauli matrices acting in the spinor space.
Thus, we can define the two-component superfields for charmed and anticharmed mesons: 
\begin{align}
    H_a = \vec{P}_{a}^{*(Q)} \cdot \vec{\sigma} + P_{a}^{(Q)}, \qquad
    \bar H^a = \vec{P}^{*(\bar Q)a} \cdot \vec{\sigma} + P^{(\bar Q)a}.
\end{align}
Then the Hermitian conjugate fields in Eq.~\eqref{eq:Hdag} become
\begin{align}
    \bar H^{(Q)a} = \begin{pmatrix}
    0 & 0 \\ H_a^\dagger & 0
    \end{pmatrix}, \qquad
    \bar H^{(\bar Q)}_a = \begin{pmatrix}
    0 & 0 \\ \bar H^{a\dagger} & 0
    \end{pmatrix}.
\end{align}
With this notation, the Lagrangian in Eq.~\eqref{eq:LaLO} reads
\begin{align}
\mathcal{L}_{4H} =&-\frac14\,\Tr\left[H_{}^{a\dagger} H_{b}^{}\right]\Tr\left[\bar{H}_{}^{c} \bar{H}_{d}^{\dagger}\right] \left( F_{A}^{}\, \delta_a^{\,b} \delta_{c}^{\,d} + F_{A}^{\lambda}\, \vec\lambda_a^{\,b} \cdot \vec\lambda_c^{\,d} \right) \notag \\
&+\frac14\, \Tr\left[H_{}^{a\dagger} H_{b}^{} \sigma^{m}\right] \Tr\left[\bar{H}_{}^{c} \bar{H}_{d}^{\dagger} \sigma^{m}\right] \left( F_{B}^{}\, \delta_a^{\,b} \delta_{c}^{\,d} + F_{B}^{\lambda}\, \vec\lambda_a^{\,b} \cdot \vec\lambda_c^{\,d} \right) .
\label{eq:L0_2component}
\end{align}
Using the completeness relations for the Pauli and Gell-Mann matrices, 
\begin{align}
    \delta_{i}^{\,l} \delta_{k}^{\,j}=\frac{1}{2} \delta_{i}^{\,j} \delta_{k}^{\,l} +\frac{1}{2}\vec\sigma_{i}^{\,j}\cdot \vec\sigma_{k}^{\,l},  \qquad 
    \delta_{a}^{\,d} \delta_{c}^{\,b}=\frac{1}{3} \delta_{a}^{\,b} \delta_{c}^{\,d} +\frac{1}{2}\vec\lambda_{a}^{\,b}\cdot\vec \lambda_{c}^{\,d},
\end{align}
one can check that the double-trace form in Eq.~\eqref{eq:L0_2component} can be rewritten in the single-trace form given in Ref.~\cite{Mehen:2011yh}, where the light-flavor SU(2) case was considered. 

\section{Error estimation}
\label{app:errors}

To estimate the uncertainties on the pole positions inherited from those of the LECs, we generate MC samplings of the physical input values according to their central values and errors, as discussed in Sect.~\ref{sect:spectrum}. Here, we show in Fig.~\ref{fig:hist} some examples  of pole-position distributions, relative to the corresponding thresholds. They have obtained using the S-I scheme with $\Lambda=0.5$ GeV, and taking into account the dispersion produced by HQSS and SU(3) flavor breaking effects. The central pole positions for these channels are obtained by using the central values of the inputs, and the asymmetric upper and lower errors are determined from 68\% confident level limits, referred to the central value. Hence these distributions give rise to the second set of uncertainties  in Tables ~\ref{tab:I=0}, \ref{tab:hiddens}, \ref{tab:I=1/2} and \ref{tab:isovector}, which we consider as  conservative estimates of the total errors on our predictions. Note that we have added an additional minus sign to the virtual pole positions for better illustration, and that one should always keep in mind the poles are always below the corresponding thresholds. This is to say, as in the tables, a negative (positive) value of $E$ means a bound (virtual) state located on the first (second) RS below threshold, and the corresponding histogram is coloured in green (orange).
 In some sectors, for instance the $(I)J^P=(1/2)0^+$ $D_s\bar D$ and $(I)J^P=(1/2)2^+$ $D_s^*\bar D^*$ channels, only a small fraction (less than 10\%) of samples of the LECs produce near-threshold poles and most samples yield repulsive or too weak attractive interactions. Therefore, we conclude that these channels are unlikely to have near-threshold molecular states.
\begin{figure}
    \centering
    \includegraphics[width=\linewidth]{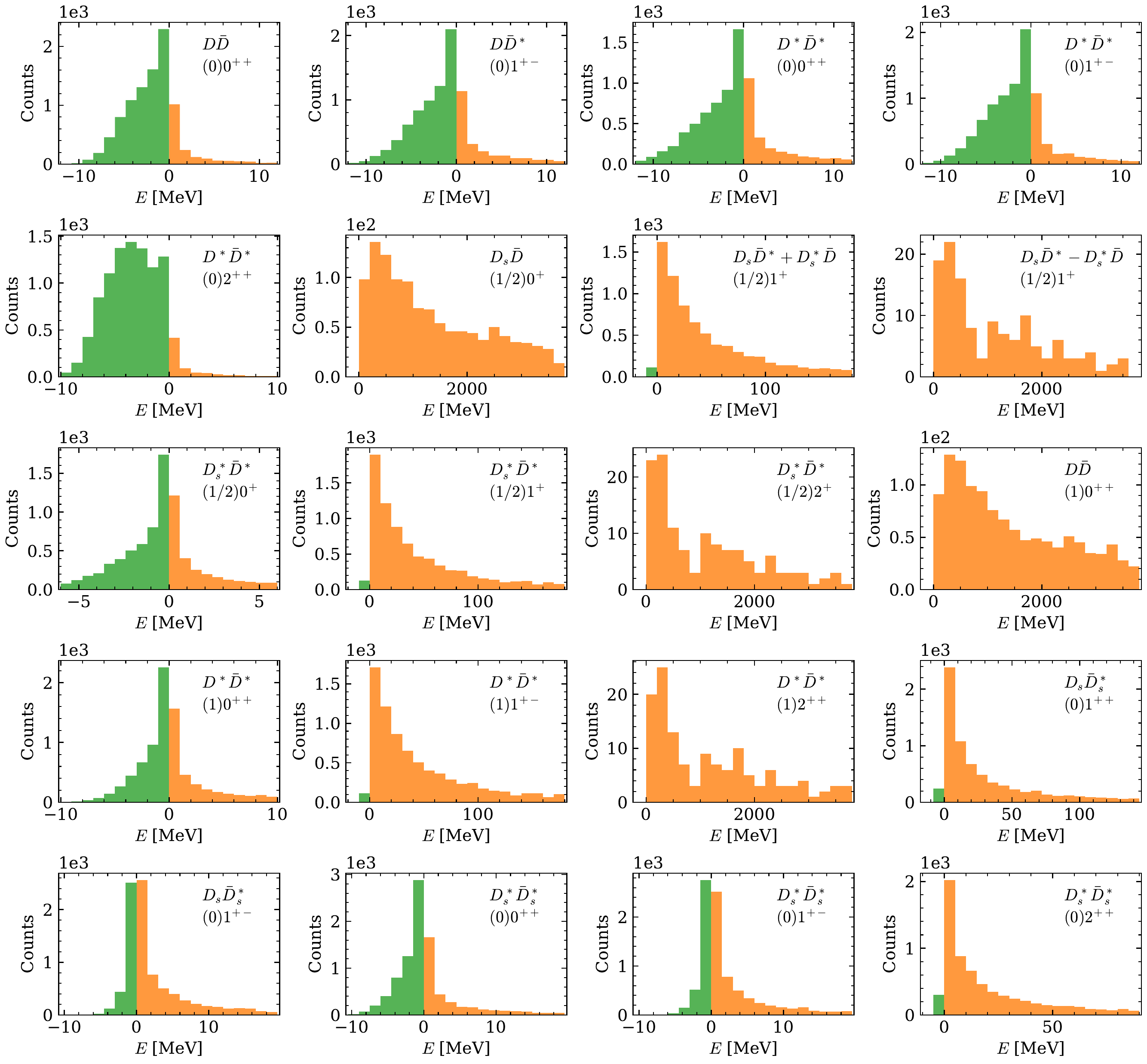}
    \caption{Pole-position histograms obtained from a $10^4$ MC sampling of the S-I LECs, including both the uncertainty from the physical inputs discussed in the text and that produced by  HQSS and SU(3) flavor symmetry breaking effects. The  cutoff has been fixed to  $\Lambda=0.5$ GeV. See text for details.}
    \label{fig:hist}
\end{figure}
\end{appendix}

\bibliography{DsDsrefs.bib}

\begin{thebibliography}{74}%
\makeatletter
\providecommand \@ifxundefined [1]{%
 \@ifx{#1\undefined}
}%
\providecommand \@ifnum [1]{%
 \ifnum #1\expandafter \@firstoftwo
 \else \expandafter \@secondoftwo
 \fi
}%
\providecommand \@ifx [1]{%
 \ifx #1\expandafter \@firstoftwo
 \else \expandafter \@secondoftwo
 \fi
}%
\providecommand \natexlab [1]{#1}%
\providecommand \enquote  [1]{``#1''}%
\providecommand \bibnamefont  [1]{#1}%
\providecommand \bibfnamefont [1]{#1}%
\providecommand \citenamefont [1]{#1}%
\providecommand \href@noop [0]{\@secondoftwo}%
\providecommand \href [0]{\begingroup \@sanitize@url \@href}%
\providecommand \@href[1]{\@@startlink{#1}\@@href}%
\providecommand \@@href[1]{\endgroup#1\@@endlink}%
\providecommand \@sanitize@url [0]{\catcode `\\12\catcode `\$12\catcode
  `\&12\catcode `\#12\catcode `\^12\catcode `\_12\catcode `\%12\relax}%
\providecommand \@@startlink[1]{}%
\providecommand \@@endlink[0]{}%
\providecommand \url  [0]{\begingroup\@sanitize@url \@url }%
\providecommand \@url [1]{\endgroup\@href {#1}{\urlprefix }}%
\providecommand \urlprefix  [0]{URL }%
\providecommand \Eprint [0]{\href }%
\providecommand \doibase [0]{https://doi.org/}%
\providecommand \selectlanguage [0]{\@gobble}%
\providecommand \bibinfo  [0]{\@secondoftwo}%
\providecommand \bibfield  [0]{\@secondoftwo}%
\providecommand \translation [1]{[#1]}%
\providecommand \BibitemOpen [0]{}%
\providecommand \bibitemStop [0]{}%
\providecommand \bibitemNoStop [0]{.\EOS\space}%
\providecommand \EOS [0]{\spacefactor3000\relax}%
\providecommand \BibitemShut  [1]{\csname bibitem#1\endcsname}%
\let\auto@bib@innerbib\@empty
\bibitem [{\citenamefont {Choi}\ \emph {et~al.}(2003)\citenamefont {Choi} \emph
  {et~al.}}]{Belle:2003nnu}%
  \BibitemOpen
  \bibfield  {author} {\bibinfo {author} {\bibfnamefont {S.~K.}\ \bibnamefont
  {Choi}} \emph {et~al.} (\bibinfo {collaboration} {Belle}),\ }\bibfield
  {title} {\bibinfo {title} {Observation of a narrow charmonium-like state in
  exclusive ${B}^{\pm} \rightarrow {K}^{\pm} \pi^{+} \pi^- {J} / \psi$
  decays},\ }\href {https://doi.org/10.1103/PhysRevLett.91.262001} {\bibfield
  {journal} {\bibinfo  {journal} {Phys. Rev. Lett.}\ }\textbf {\bibinfo
  {volume} {91}},\ \bibinfo {pages} {262001} (\bibinfo {year} {2003})},\
  \Eprint {https://arxiv.org/abs/hep-ex/0309032} {arXiv:hep-ex/0309032}
  \BibitemShut {NoStop}%
\bibitem [{\citenamefont {Ablikim}\ \emph
  {et~al.}(2013{\natexlab{a}})\citenamefont {Ablikim} \emph
  {et~al.}}]{BESIII:2013ris}%
  \BibitemOpen
  \bibfield  {author} {\bibinfo {author} {\bibfnamefont {M.}~\bibnamefont
  {Ablikim}} \emph {et~al.} (\bibinfo {collaboration} {BESIII}),\ }\bibfield
  {title} {\bibinfo {title} {Observation of a charged charmoniumlike structure
  in {{$e^+e^-\to \pi^{+}\pi^-J/\psi$}} at {${\sqrt{s} =4.26}$} {{GeV}}},\
  }\href {https://doi.org/10.1103/PhysRevLett.110.252001} {\bibfield  {journal}
  {\bibinfo  {journal} {Phys. Rev. Lett.}\ }\textbf {\bibinfo {volume} {110}},\
  \bibinfo {pages} {252001} (\bibinfo {year} {2013}{\natexlab{a}})},\ \Eprint
  {https://arxiv.org/abs/1303.5949} {arXiv:1303.5949 [hep-ex]} \BibitemShut
  {NoStop}%
\bibitem [{\citenamefont {Liu}\ \emph {et~al.}(2013)\citenamefont {Liu} \emph
  {et~al.}}]{Belle:2013yex}%
  \BibitemOpen
  \bibfield  {author} {\bibinfo {author} {\bibfnamefont {Z.~Q.}\ \bibnamefont
  {Liu}} \emph {et~al.} (\bibinfo {collaboration} {Belle}),\ }\bibfield
  {title} {\bibinfo {title} {Study of {{$e^+e^-\to \pi^{+}\pi^-J/\psi$}} and
  observation of a charged charmoniumlike state at belle},\ }\href
  {https://doi.org/10.1103/PhysRevLett.110.252002} {\bibfield  {journal}
  {\bibinfo  {journal} {Phys. Rev. Lett.}\ }\textbf {\bibinfo {volume} {110}},\
  \bibinfo {pages} {252002} (\bibinfo {year} {2013})},\ \Eprint
  {https://arxiv.org/abs/1304.0121} {arXiv:1304.0121 [hep-ex]} \BibitemShut
  {NoStop}%
\bibitem [{\citenamefont {Barnes}\ and\ \citenamefont
  {Godfrey}(2004)}]{Barnes:2003vb}%
  \BibitemOpen
  \bibfield  {author} {\bibinfo {author} {\bibfnamefont {T.}~\bibnamefont
  {Barnes}}\ and\ \bibinfo {author} {\bibfnamefont {S.}~\bibnamefont
  {Godfrey}},\ }\bibfield  {title} {\bibinfo {title} {{Charmonium options for
  the $X(3872)$}},\ }\href {https://doi.org/10.1103/PhysRevD.69.054008}
  {\bibfield  {journal} {\bibinfo  {journal} {Phys. Rev. D}\ }\textbf {\bibinfo
  {volume} {69}},\ \bibinfo {pages} {054008} (\bibinfo {year} {2004})},\
  \Eprint {https://arxiv.org/abs/hep-ph/0311162} {arXiv:hep-ph/0311162}
  \BibitemShut {NoStop}%
\bibitem [{\citenamefont {Voloshin}(2004)}]{Voloshin:2003nt}%
  \BibitemOpen
  \bibfield  {author} {\bibinfo {author} {\bibfnamefont {M.~B.}\ \bibnamefont
  {Voloshin}},\ }\bibfield  {title} {\bibinfo {title} {{Interference and
  binding effects in decays of possible molecular component of $X(3872)$}},\
  }\href {https://doi.org/10.1016/j.physletb.2003.11.014} {\bibfield  {journal}
  {\bibinfo  {journal} {Phys. Lett. B}\ }\textbf {\bibinfo {volume} {579}},\
  \bibinfo {pages} {316} (\bibinfo {year} {2004})},\ \Eprint
  {https://arxiv.org/abs/hep-ph/0309307} {arXiv:hep-ph/0309307} \BibitemShut
  {NoStop}%
\bibitem [{\citenamefont {Swanson}(2004{\natexlab{a}})}]{Swanson:2003tb}%
  \BibitemOpen
  \bibfield  {author} {\bibinfo {author} {\bibfnamefont {E.~S.}\ \bibnamefont
  {Swanson}},\ }\bibfield  {title} {\bibinfo {title} {{Short range structure in
  the $X(3872)$}},\ }\href {https://doi.org/10.1016/j.physletb.2004.03.033}
  {\bibfield  {journal} {\bibinfo  {journal} {Phys. Lett. B}\ }\textbf
  {\bibinfo {volume} {588}},\ \bibinfo {pages} {189} (\bibinfo {year}
  {2004}{\natexlab{a}})},\ \Eprint {https://arxiv.org/abs/hep-ph/0311229}
  {arXiv:hep-ph/0311229} \BibitemShut {NoStop}%
\bibitem [{\citenamefont {T{\"o}rnqvist}(2004)}]{Tornqvist:2004qy}%
  \BibitemOpen
  \bibfield  {author} {\bibinfo {author} {\bibfnamefont {N.~A.}\ \bibnamefont
  {T{\"o}rnqvist}},\ }\bibfield  {title} {\bibinfo {title} {{Isospin breaking
  of the narrow charmonium state of Belle at 3872 MeV as a deuson}},\ }\href
  {https://doi.org/10.1016/j.physletb.2004.03.077} {\bibfield  {journal}
  {\bibinfo  {journal} {Phys. Lett. B}\ }\textbf {\bibinfo {volume} {590}},\
  \bibinfo {pages} {209} (\bibinfo {year} {2004})},\ \Eprint
  {https://arxiv.org/abs/hep-ph/0402237} {arXiv:hep-ph/0402237} \BibitemShut
  {NoStop}%
\bibitem [{\citenamefont {Swanson}(2004{\natexlab{b}})}]{Swanson:2004pp}%
  \BibitemOpen
  \bibfield  {author} {\bibinfo {author} {\bibfnamefont {E.~S.}\ \bibnamefont
  {Swanson}},\ }\bibfield  {title} {\bibinfo {title} {{Diagnostic decays of the
  $X(3872)$}},\ }\href {https://doi.org/10.1016/j.physletb.2004.07.059}
  {\bibfield  {journal} {\bibinfo  {journal} {Phys. Lett. B}\ }\textbf
  {\bibinfo {volume} {598}},\ \bibinfo {pages} {197} (\bibinfo {year}
  {2004}{\natexlab{b}})},\ \Eprint {https://arxiv.org/abs/hep-ph/0406080}
  {arXiv:hep-ph/0406080} \BibitemShut {NoStop}%
\bibitem [{\citenamefont {Suzuki}(2005)}]{Suzuki:2005ha}%
  \BibitemOpen
  \bibfield  {author} {\bibinfo {author} {\bibfnamefont {M.}~\bibnamefont
  {Suzuki}},\ }\bibfield  {title} {\bibinfo {title} {{The $X(3872)$ boson:
  Molecule or charmonium}},\ }\href
  {https://doi.org/10.1103/PhysRevD.72.114013} {\bibfield  {journal} {\bibinfo
  {journal} {Phys. Rev. D}\ }\textbf {\bibinfo {volume} {72}},\ \bibinfo
  {pages} {114013} (\bibinfo {year} {2005})},\ \Eprint
  {https://arxiv.org/abs/hep-ph/0508258} {arXiv:hep-ph/0508258} \BibitemShut
  {NoStop}%
\bibitem [{\citenamefont {AlFiky}\ \emph {et~al.}(2006)\citenamefont {AlFiky},
  \citenamefont {Gabbiani},\ and\ \citenamefont {Petrov}}]{AlFiky:2005jd}%
  \BibitemOpen
  \bibfield  {author} {\bibinfo {author} {\bibfnamefont {M.~T.}\ \bibnamefont
  {AlFiky}}, \bibinfo {author} {\bibfnamefont {F.}~\bibnamefont {Gabbiani}},\
  and\ \bibinfo {author} {\bibfnamefont {A.~A.}\ \bibnamefont {Petrov}},\
  }\bibfield  {title} {\bibinfo {title} {{$X(3872)$: {{Hadronic}} molecules in
  effective field theory}},\ }\href
  {https://doi.org/10.1016/j.physletb.2006.07.069} {\bibfield  {journal}
  {\bibinfo  {journal} {Phys. Lett. B}\ }\textbf {\bibinfo {volume} {640}},\
  \bibinfo {pages} {238} (\bibinfo {year} {2006})},\ \Eprint
  {https://arxiv.org/abs/hep-ph/0506141} {arXiv:hep-ph/0506141} \BibitemShut
  {NoStop}%
\bibitem [{\citenamefont {Gamermann}\ \emph {et~al.}(2007)\citenamefont
  {Gamermann}, \citenamefont {Oset}, \citenamefont {Strottman},\ and\
  \citenamefont {Vicente~Vacas}}]{Gamermann:2006nm}%
  \BibitemOpen
  \bibfield  {author} {\bibinfo {author} {\bibfnamefont {D.}~\bibnamefont
  {Gamermann}}, \bibinfo {author} {\bibfnamefont {E.}~\bibnamefont {Oset}},
  \bibinfo {author} {\bibfnamefont {D.}~\bibnamefont {Strottman}},\ and\
  \bibinfo {author} {\bibfnamefont {M.~J.}\ \bibnamefont {Vicente~Vacas}},\
  }\bibfield  {title} {\bibinfo {title} {Dynamically generated open and hidden
  charm meson systems},\ }\href {https://doi.org/10.1103/PhysRevD.76.074016}
  {\bibfield  {journal} {\bibinfo  {journal} {Phys. Rev. D}\ }\textbf {\bibinfo
  {volume} {76}},\ \bibinfo {pages} {074016} (\bibinfo {year} {2007})},\
  \Eprint {https://arxiv.org/abs/hep-ph/0612179} {arXiv:hep-ph/0612179}
  \BibitemShut {NoStop}%
\bibitem [{\citenamefont {Hanhart}\ \emph {et~al.}(2007)\citenamefont
  {Hanhart}, \citenamefont {Kalashnikova}, \citenamefont {Kudryavtsev},\ and\
  \citenamefont {Nefediev}}]{Hanhart:2007yq}%
  \BibitemOpen
  \bibfield  {author} {\bibinfo {author} {\bibfnamefont {C.}~\bibnamefont
  {Hanhart}}, \bibinfo {author} {\bibfnamefont {Y.~S.}\ \bibnamefont
  {Kalashnikova}}, \bibinfo {author} {\bibfnamefont {A.~E.}\ \bibnamefont
  {Kudryavtsev}},\ and\ \bibinfo {author} {\bibfnamefont {A.~V.}\ \bibnamefont
  {Nefediev}},\ }\bibfield  {title} {\bibinfo {title} {{Reconciling the
  $X(3872)$ with the near-threshold enhancement in the $D^0 \bar D^{*0}$ final
  state}},\ }\href {https://doi.org/10.1103/PhysRevD.76.034007} {\bibfield
  {journal} {\bibinfo  {journal} {Phys. Rev. D}\ }\textbf {\bibinfo {volume}
  {76}},\ \bibinfo {pages} {034007} (\bibinfo {year} {2007})},\ \Eprint
  {https://arxiv.org/abs/0704.0605} {arXiv:0704.0605 [hep-ph]} \BibitemShut
  {NoStop}%
\bibitem [{\citenamefont {Fleming}\ \emph {et~al.}(2007)\citenamefont
  {Fleming}, \citenamefont {Kusunoki}, \citenamefont {Mehen},\ and\
  \citenamefont {van Kolck}}]{Fleming:2007rp}%
  \BibitemOpen
  \bibfield  {author} {\bibinfo {author} {\bibfnamefont {S.}~\bibnamefont
  {Fleming}}, \bibinfo {author} {\bibfnamefont {M.}~\bibnamefont {Kusunoki}},
  \bibinfo {author} {\bibfnamefont {T.}~\bibnamefont {Mehen}},\ and\ \bibinfo
  {author} {\bibfnamefont {U.}~\bibnamefont {van Kolck}},\ }\bibfield  {title}
  {\bibinfo {title} {{Pion interactions in the $X(3872)$}},\ }\href
  {https://doi.org/10.1103/PhysRevD.76.034006} {\bibfield  {journal} {\bibinfo
  {journal} {Phys. Rev. D}\ }\textbf {\bibinfo {volume} {76}},\ \bibinfo
  {pages} {034006} (\bibinfo {year} {2007})},\ \Eprint
  {https://arxiv.org/abs/hep-ph/0703168} {arXiv:hep-ph/0703168} \BibitemShut
  {NoStop}%
\bibitem [{\citenamefont {Braaten}\ and\ \citenamefont
  {Lu}(2007)}]{Braaten:2007dw}%
  \BibitemOpen
  \bibfield  {author} {\bibinfo {author} {\bibfnamefont {E.}~\bibnamefont
  {Braaten}}\ and\ \bibinfo {author} {\bibfnamefont {M.}~\bibnamefont {Lu}},\
  }\bibfield  {title} {\bibinfo {title} {{Line shapes of the $X(3872)$}},\
  }\href {https://doi.org/10.1103/PhysRevD.76.094028} {\bibfield  {journal}
  {\bibinfo  {journal} {Phys. Rev. D}\ }\textbf {\bibinfo {volume} {76}},\
  \bibinfo {pages} {094028} (\bibinfo {year} {2007})},\ \Eprint
  {https://arxiv.org/abs/0709.2697} {arXiv:0709.2697 [hep-ph]} \BibitemShut
  {NoStop}%
\bibitem [{\citenamefont {Thomas}\ and\ \citenamefont
  {Close}(2008)}]{Thomas:2008ja}%
  \BibitemOpen
  \bibfield  {author} {\bibinfo {author} {\bibfnamefont {C.~E.}\ \bibnamefont
  {Thomas}}\ and\ \bibinfo {author} {\bibfnamefont {F.~E.}\ \bibnamefont
  {Close}},\ }\bibfield  {title} {\bibinfo {title} {{Is $X(3872)$ a
  molecule?}},\ }\href {https://doi.org/10.1103/PhysRevD.78.034007} {\bibfield
  {journal} {\bibinfo  {journal} {Phys. Rev. D}\ }\textbf {\bibinfo {volume}
  {78}},\ \bibinfo {pages} {034007} (\bibinfo {year} {2008})},\ \Eprint
  {https://arxiv.org/abs/0805.3653} {arXiv:0805.3653 [hep-ph]} \BibitemShut
  {NoStop}%
\bibitem [{\citenamefont {Fleming}\ and\ \citenamefont
  {Mehen}(2008)}]{Fleming:2008yn}%
  \BibitemOpen
  \bibfield  {author} {\bibinfo {author} {\bibfnamefont {S.}~\bibnamefont
  {Fleming}}\ and\ \bibinfo {author} {\bibfnamefont {T.}~\bibnamefont
  {Mehen}},\ }\bibfield  {title} {\bibinfo {title} {{Hadronic decays of the
  $X(3872)$ to $\chi_{cJ}$ in effective field theory}},\ }\href
  {https://doi.org/10.1103/PhysRevD.78.094019} {\bibfield  {journal} {\bibinfo
  {journal} {Phys. Rev. D}\ }\textbf {\bibinfo {volume} {78}},\ \bibinfo
  {pages} {094019} (\bibinfo {year} {2008})},\ \Eprint
  {https://arxiv.org/abs/0807.2674} {arXiv:0807.2674 [hep-ph]} \BibitemShut
  {NoStop}%
\bibitem [{\citenamefont {Liu}\ \emph {et~al.}(2008)\citenamefont {Liu},
  \citenamefont {Liu}, \citenamefont {Deng},\ and\ \citenamefont
  {Zhu}}]{Liu:2008fh}%
  \BibitemOpen
  \bibfield  {author} {\bibinfo {author} {\bibfnamefont {Y.-R.}\ \bibnamefont
  {Liu}}, \bibinfo {author} {\bibfnamefont {X.}~\bibnamefont {Liu}}, \bibinfo
  {author} {\bibfnamefont {W.-Z.}\ \bibnamefont {Deng}},\ and\ \bibinfo
  {author} {\bibfnamefont {S.-L.}\ \bibnamefont {Zhu}},\ }\bibfield  {title}
  {\bibinfo {title} {{Is $X(3872) $ really a molecular state?}},\ }\href
  {https://doi.org/10.1140/epjc/s10052-008-0640-4} {\bibfield  {journal}
  {\bibinfo  {journal} {Eur. Phys. J. C}\ }\textbf {\bibinfo {volume} {56}},\
  \bibinfo {pages} {63} (\bibinfo {year} {2008})},\ \Eprint
  {https://arxiv.org/abs/0801.3540} {arXiv:0801.3540 [hep-ph]} \BibitemShut
  {NoStop}%
\bibitem [{\citenamefont {Liu}\ \emph {et~al.}(2009)\citenamefont {Liu},
  \citenamefont {Luo}, \citenamefont {Liu},\ and\ \citenamefont
  {Zhu}}]{Liu:2009qhy}%
  \BibitemOpen
  \bibfield  {author} {\bibinfo {author} {\bibfnamefont {X.}~\bibnamefont
  {Liu}}, \bibinfo {author} {\bibfnamefont {Z.-G.}\ \bibnamefont {Luo}},
  \bibinfo {author} {\bibfnamefont {Y.-R.}\ \bibnamefont {Liu}},\ and\ \bibinfo
  {author} {\bibfnamefont {S.-L.}\ \bibnamefont {Zhu}},\ }\bibfield  {title}
  {\bibinfo {title} {{$X(3872)$ and other possible heavy molecular states}},\
  }\href {https://doi.org/10.1140/epjc/s10052-009-1020-4} {\bibfield  {journal}
  {\bibinfo  {journal} {Eur. Phys. J. C}\ }\textbf {\bibinfo {volume} {61}},\
  \bibinfo {pages} {411} (\bibinfo {year} {2009})},\ \Eprint
  {https://arxiv.org/abs/0808.0073} {arXiv:0808.0073 [hep-ph]} \BibitemShut
  {NoStop}%
\bibitem [{\citenamefont {Gamermann}\ and\ \citenamefont
  {Oset}(2009)}]{Gamermann:2009fv}%
  \BibitemOpen
  \bibfield  {author} {\bibinfo {author} {\bibfnamefont {D.}~\bibnamefont
  {Gamermann}}\ and\ \bibinfo {author} {\bibfnamefont {E.}~\bibnamefont
  {Oset}},\ }\bibfield  {title} {\bibinfo {title} {{Isospin breaking effects in
  the $X(3872)$ resonance}},\ }\href
  {https://doi.org/10.1103/PhysRevD.80.014003} {\bibfield  {journal} {\bibinfo
  {journal} {Phys. Rev. D}\ }\textbf {\bibinfo {volume} {80}},\ \bibinfo
  {pages} {014003} (\bibinfo {year} {2009})},\ \Eprint
  {https://arxiv.org/abs/0905.0402} {arXiv:0905.0402 [hep-ph]} \BibitemShut
  {NoStop}%
\bibitem [{\citenamefont {Gamermann}\ \emph {et~al.}(2010)\citenamefont
  {Gamermann}, \citenamefont {Nieves}, \citenamefont {Oset},\ and\
  \citenamefont {Ruiz~Arriola}}]{Gamermann:2009uq}%
  \BibitemOpen
  \bibfield  {author} {\bibinfo {author} {\bibfnamefont {D.}~\bibnamefont
  {Gamermann}}, \bibinfo {author} {\bibfnamefont {J.}~\bibnamefont {Nieves}},
  \bibinfo {author} {\bibfnamefont {E.}~\bibnamefont {Oset}},\ and\ \bibinfo
  {author} {\bibfnamefont {E.}~\bibnamefont {Ruiz~Arriola}},\ }\bibfield
  {title} {\bibinfo {title} {{Couplings in coupled channels versus wave
  functions: application to the $X(3872)$ resonance}},\ }\href
  {https://doi.org/10.1103/PhysRevD.81.014029} {\bibfield  {journal} {\bibinfo
  {journal} {Phys. Rev. D}\ }\textbf {\bibinfo {volume} {81}},\ \bibinfo
  {pages} {014029} (\bibinfo {year} {2010})},\ \Eprint
  {https://arxiv.org/abs/0911.4407} {arXiv:0911.4407 [hep-ph]} \BibitemShut
  {NoStop}%
\bibitem [{\citenamefont {Bignamini}\ \emph {et~al.}(2009)\citenamefont
  {Bignamini}, \citenamefont {Grinstein}, \citenamefont {Piccinini},
  \citenamefont {Polosa},\ and\ \citenamefont {Sabelli}}]{Bignamini:2009sk}%
  \BibitemOpen
  \bibfield  {author} {\bibinfo {author} {\bibfnamefont {C.}~\bibnamefont
  {Bignamini}}, \bibinfo {author} {\bibfnamefont {B.}~\bibnamefont
  {Grinstein}}, \bibinfo {author} {\bibfnamefont {F.}~\bibnamefont
  {Piccinini}}, \bibinfo {author} {\bibfnamefont {A.~D.}\ \bibnamefont
  {Polosa}},\ and\ \bibinfo {author} {\bibfnamefont {C.}~\bibnamefont
  {Sabelli}},\ }\bibfield  {title} {\bibinfo {title} {{Is the $X(3872)$
  production cross section at Tevatron compatible with a hadron molecule
  interpretation?}},\ }\href {https://doi.org/10.1103/PhysRevLett.103.162001}
  {\bibfield  {journal} {\bibinfo  {journal} {Phys. Rev. Lett.}\ }\textbf
  {\bibinfo {volume} {103}},\ \bibinfo {pages} {162001} (\bibinfo {year}
  {2009})},\ \Eprint {https://arxiv.org/abs/0906.0882} {arXiv:0906.0882
  [hep-ph]} \BibitemShut {NoStop}%
\bibitem [{\citenamefont {Nieves}\ and\ \citenamefont
  {Valderrama}(2011)}]{Nieves:2011zz}%
  \BibitemOpen
  \bibfield  {author} {\bibinfo {author} {\bibfnamefont {J.}~\bibnamefont
  {Nieves}}\ and\ \bibinfo {author} {\bibfnamefont {M.~P.}\ \bibnamefont
  {Valderrama}},\ }\bibfield  {title} {\bibinfo {title} {{Deriving the
  existence of $B\bar{B}^*$ bound states from the $X(3872)$ and heavy quark
  symmetry}},\ }\href {https://doi.org/10.1103/PhysRevD.84.056015} {\bibfield
  {journal} {\bibinfo  {journal} {Phys. Rev. D}\ }\textbf {\bibinfo {volume}
  {84}},\ \bibinfo {pages} {056015} (\bibinfo {year} {2011})},\ \Eprint
  {https://arxiv.org/abs/1106.0600} {arXiv:1106.0600 [hep-ph]} \BibitemShut
  {NoStop}%
\bibitem [{\citenamefont {Nieves}\ and\ \citenamefont
  {Valderrama}(2012)}]{Nieves:2012tt}%
  \BibitemOpen
  \bibfield  {author} {\bibinfo {author} {\bibfnamefont {J.}~\bibnamefont
  {Nieves}}\ and\ \bibinfo {author} {\bibfnamefont {M.~P.}\ \bibnamefont
  {Valderrama}},\ }\bibfield  {title} {\bibinfo {title} {The heavy quark spin
  symmetry partners of the {{$X$(3872)}}},\ }\href
  {https://doi.org/10.1103/PhysRevD.86.056004} {\bibfield  {journal} {\bibinfo
  {journal} {Phys. Rev. D}\ }\textbf {\bibinfo {volume} {86}},\ \bibinfo
  {pages} {056004} (\bibinfo {year} {2012})},\ \Eprint
  {https://arxiv.org/abs/1204.2790} {arXiv:1204.2790 [hep-ph]} \BibitemShut
  {NoStop}%
\bibitem [{\citenamefont {Guo}\ \emph {et~al.}(2013{\natexlab{a}})\citenamefont
  {Guo}, \citenamefont {Hanhart}, \citenamefont {Mei\ss{}ner}, \citenamefont
  {Wang},\ and\ \citenamefont {Zhao}}]{Guo:2013zbw}%
  \BibitemOpen
  \bibfield  {author} {\bibinfo {author} {\bibfnamefont {F.-K.}\ \bibnamefont
  {Guo}}, \bibinfo {author} {\bibfnamefont {C.}~\bibnamefont {Hanhart}},
  \bibinfo {author} {\bibfnamefont {U.-G.}\ \bibnamefont {Mei\ss{}ner}},
  \bibinfo {author} {\bibfnamefont {Q.}~\bibnamefont {Wang}},\ and\ \bibinfo
  {author} {\bibfnamefont {Q.}~\bibnamefont {Zhao}},\ }\bibfield  {title}
  {\bibinfo {title} {{Production of the $X(3872)$ in charmonia radiative
  decays}},\ }\href {https://doi.org/10.1016/j.physletb.2013.06.053} {\bibfield
   {journal} {\bibinfo  {journal} {Phys. Lett. B}\ }\textbf {\bibinfo {volume}
  {725}},\ \bibinfo {pages} {127} (\bibinfo {year} {2013}{\natexlab{a}})},\
  \Eprint {https://arxiv.org/abs/1306.3096} {arXiv:1306.3096 [hep-ph]}
  \BibitemShut {NoStop}%
\bibitem [{\citenamefont {Voloshin}(2013)}]{Voloshin:2013dpa}%
  \BibitemOpen
  \bibfield  {author} {\bibinfo {author} {\bibfnamefont {M.~B.}\ \bibnamefont
  {Voloshin}},\ }\bibfield  {title} {\bibinfo {title} {{$Z_c(3900)$ -- what is
  inside?}},\ }\href {https://doi.org/10.1103/PhysRevD.87.091501} {\bibfield
  {journal} {\bibinfo  {journal} {Phys. Rev. D}\ }\textbf {\bibinfo {volume}
  {87}},\ \bibinfo {pages} {091501} (\bibinfo {year} {2013})},\ \Eprint
  {https://arxiv.org/abs/1304.0380} {arXiv:1304.0380 [hep-ph]} \BibitemShut
  {NoStop}%
\bibitem [{\citenamefont {Wang}\ and\ \citenamefont
  {Huang}(2014)}]{Wang:2013vex}%
  \BibitemOpen
  \bibfield  {author} {\bibinfo {author} {\bibfnamefont {Z.-G.}\ \bibnamefont
  {Wang}}\ and\ \bibinfo {author} {\bibfnamefont {T.}~\bibnamefont {Huang}},\
  }\bibfield  {title} {\bibinfo {title} {{Analysis of the $X(3872)$,
  $Z_c(3900)$ and $Z_c(3885)$ as axial-vector tetraquark states with QCD sum
  rules}},\ }\href {https://doi.org/10.1103/PhysRevD.89.054019} {\bibfield
  {journal} {\bibinfo  {journal} {Phys. Rev. D}\ }\textbf {\bibinfo {volume}
  {89}},\ \bibinfo {pages} {054019} (\bibinfo {year} {2014})},\ \Eprint
  {https://arxiv.org/abs/1310.2422} {arXiv:1310.2422 [hep-ph]} \BibitemShut
  {NoStop}%
\bibitem [{\citenamefont {{Hidalgo-Duque}}\ \emph
  {et~al.}(2013{\natexlab{a}})\citenamefont {{Hidalgo-Duque}}, \citenamefont
  {Nieves},\ and\ \citenamefont {Valderrama}}]{Hidalgo-Duque:2012rqv}%
  \BibitemOpen
  \bibfield  {author} {\bibinfo {author} {\bibfnamefont {C.}~\bibnamefont
  {{Hidalgo-Duque}}}, \bibinfo {author} {\bibfnamefont {J.}~\bibnamefont
  {Nieves}},\ and\ \bibinfo {author} {\bibfnamefont {M.~P.}\ \bibnamefont
  {Valderrama}},\ }\bibfield  {title} {\bibinfo {title} {Light flavor and heavy
  quark spin symmetry in heavy meson molecules},\ }\href
  {https://doi.org/10.1103/PhysRevD.87.076006} {\bibfield  {journal} {\bibinfo
  {journal} {Phys. Rev. D}\ }\textbf {\bibinfo {volume} {87}},\ \bibinfo
  {pages} {076006} (\bibinfo {year} {2013}{\natexlab{a}})},\ \Eprint
  {https://arxiv.org/abs/1210.5431} {arXiv:1210.5431 [hep-ph]} \BibitemShut
  {NoStop}%
\bibitem [{\citenamefont {Prelovsek}\ and\ \citenamefont
  {Leskovec}(2013)}]{Prelovsek:2013cra}%
  \BibitemOpen
  \bibfield  {author} {\bibinfo {author} {\bibfnamefont {S.}~\bibnamefont
  {Prelovsek}}\ and\ \bibinfo {author} {\bibfnamefont {L.}~\bibnamefont
  {Leskovec}},\ }\bibfield  {title} {\bibinfo {title} {{Evidence for $X(3872)$
  from $DD^*$ scattering on the lattice}},\ }\href
  {https://doi.org/10.1103/PhysRevLett.111.192001} {\bibfield  {journal}
  {\bibinfo  {journal} {Phys. Rev. Lett.}\ }\textbf {\bibinfo {volume} {111}},\
  \bibinfo {pages} {192001} (\bibinfo {year} {2013})},\ \Eprint
  {https://arxiv.org/abs/1307.5172} {arXiv:1307.5172 [hep-lat]} \BibitemShut
  {NoStop}%
\bibitem [{\citenamefont {Guo}\ \emph {et~al.}(2013{\natexlab{b}})\citenamefont
  {Guo}, \citenamefont {{Hidalgo-Duque}}, \citenamefont {Nieves},\ and\
  \citenamefont {Valderrama}}]{Guo:2013sya}%
  \BibitemOpen
  \bibfield  {author} {\bibinfo {author} {\bibfnamefont {F.-K.}\ \bibnamefont
  {Guo}}, \bibinfo {author} {\bibfnamefont {C.}~\bibnamefont
  {{Hidalgo-Duque}}}, \bibinfo {author} {\bibfnamefont {J.}~\bibnamefont
  {Nieves}},\ and\ \bibinfo {author} {\bibfnamefont {M.~P.}\ \bibnamefont
  {Valderrama}},\ }\bibfield  {title} {\bibinfo {title} {Consequences of heavy
  quark symmetries for hadronic molecules},\ }\href
  {https://doi.org/10.1103/PhysRevD.88.054007} {\bibfield  {journal} {\bibinfo
  {journal} {Phys. Rev. D}\ }\textbf {\bibinfo {volume} {88}},\ \bibinfo
  {pages} {054007} (\bibinfo {year} {2013}{\natexlab{b}})},\ \Eprint
  {https://arxiv.org/abs/1303.6608} {arXiv:1303.6608 [hep-ph]} \BibitemShut
  {NoStop}%
\bibitem [{\citenamefont {{Hidalgo-Duque}}\ \emph
  {et~al.}(2013{\natexlab{b}})\citenamefont {{Hidalgo-Duque}}, \citenamefont
  {Nieves}, \citenamefont {Ozpineci},\ and\ \citenamefont
  {Zamiralov}}]{Hidalgo-Duque:2013pva}%
  \BibitemOpen
  \bibfield  {author} {\bibinfo {author} {\bibfnamefont {C.}~\bibnamefont
  {{Hidalgo-Duque}}}, \bibinfo {author} {\bibfnamefont {J.}~\bibnamefont
  {Nieves}}, \bibinfo {author} {\bibfnamefont {A.}~\bibnamefont {Ozpineci}},\
  and\ \bibinfo {author} {\bibfnamefont {V.}~\bibnamefont {Zamiralov}},\
  }\bibfield  {title} {\bibinfo {title} {{$X(3872)$ and its {{partners}} in the
  {{heavy quark limit}} of {{QCD}}}},\ }\href
  {https://doi.org/10.1016/j.physletb.2013.10.056} {\bibfield  {journal}
  {\bibinfo  {journal} {Phys. Lett. B}\ }\textbf {\bibinfo {volume} {727}},\
  \bibinfo {pages} {432} (\bibinfo {year} {2013}{\natexlab{b}})},\ \Eprint
  {https://arxiv.org/abs/1305.4487} {arXiv:1305.4487 [hep-ph]} \BibitemShut
  {NoStop}%
\bibitem [{\citenamefont {Guo}\ \emph {et~al.}(2015{\natexlab{a}})\citenamefont
  {Guo}, \citenamefont {Hanhart}, \citenamefont {Wang},\ and\ \citenamefont
  {Zhao}}]{Guo:2014iya}%
  \BibitemOpen
  \bibfield  {author} {\bibinfo {author} {\bibfnamefont {F.-K.}\ \bibnamefont
  {Guo}}, \bibinfo {author} {\bibfnamefont {C.}~\bibnamefont {Hanhart}},
  \bibinfo {author} {\bibfnamefont {Q.}~\bibnamefont {Wang}},\ and\ \bibinfo
  {author} {\bibfnamefont {Q.}~\bibnamefont {Zhao}},\ }\bibfield  {title}
  {\bibinfo {title} {{Could the near-threshold $XYZ$ states be simply kinematic
  effects?}},\ }\href {https://doi.org/10.1103/PhysRevD.91.051504} {\bibfield
  {journal} {\bibinfo  {journal} {Phys. Rev. D}\ }\textbf {\bibinfo {volume}
  {91}},\ \bibinfo {pages} {051504} (\bibinfo {year} {2015}{\natexlab{a}})},\
  \Eprint {https://arxiv.org/abs/1411.5584} {arXiv:1411.5584 [hep-ph]}
  \BibitemShut {NoStop}%
\bibitem [{\citenamefont {Guo}\ \emph {et~al.}(2014)\citenamefont {Guo},
  \citenamefont {Hidalgo-Duque}, \citenamefont {Nieves}, \citenamefont
  {Ozpineci},\ and\ \citenamefont {Valderrama}}]{Guo:2014hqa}%
  \BibitemOpen
  \bibfield  {author} {\bibinfo {author} {\bibfnamefont {F.~K.}\ \bibnamefont
  {Guo}}, \bibinfo {author} {\bibfnamefont {C.}~\bibnamefont {Hidalgo-Duque}},
  \bibinfo {author} {\bibfnamefont {J.}~\bibnamefont {Nieves}}, \bibinfo
  {author} {\bibfnamefont {A.}~\bibnamefont {Ozpineci}},\ and\ \bibinfo
  {author} {\bibfnamefont {M.~P.}\ \bibnamefont {Valderrama}},\ }\bibfield
  {title} {\bibinfo {title} {{Detecting the long-distance structure of the
  $X$(3872)}},\ }\href {https://doi.org/10.1140/epjc/s10052-014-2885-4}
  {\bibfield  {journal} {\bibinfo  {journal} {Eur. Phys. J. C}\ }\textbf
  {\bibinfo {volume} {74}},\ \bibinfo {pages} {2885} (\bibinfo {year}
  {2014})},\ \Eprint {https://arxiv.org/abs/1404.1776} {arXiv:1404.1776
  [hep-ph]} \BibitemShut {NoStop}%
\bibitem [{\citenamefont {Guo}\ \emph {et~al.}(2015{\natexlab{b}})\citenamefont
  {Guo}, \citenamefont {Hanhart}, \citenamefont {Kalashnikova}, \citenamefont
  {Mei\ss{}ner},\ and\ \citenamefont {Nefediev}}]{Guo:2014taa}%
  \BibitemOpen
  \bibfield  {author} {\bibinfo {author} {\bibfnamefont {F.-K.}\ \bibnamefont
  {Guo}}, \bibinfo {author} {\bibfnamefont {C.}~\bibnamefont {Hanhart}},
  \bibinfo {author} {\bibfnamefont {Y.~S.}\ \bibnamefont {Kalashnikova}},
  \bibinfo {author} {\bibfnamefont {U.-G.}\ \bibnamefont {Mei\ss{}ner}},\ and\
  \bibinfo {author} {\bibfnamefont {A.~V.}\ \bibnamefont {Nefediev}},\
  }\bibfield  {title} {\bibinfo {title} {{What can radiative decays of the
  $X(3872)$ teach us about its nature?}},\ }\href
  {https://doi.org/10.1016/j.physletb.2015.02.013} {\bibfield  {journal}
  {\bibinfo  {journal} {Phys. Lett. B}\ }\textbf {\bibinfo {volume} {742}},\
  \bibinfo {pages} {394} (\bibinfo {year} {2015}{\natexlab{b}})},\ \Eprint
  {https://arxiv.org/abs/1410.6712} {arXiv:1410.6712 [hep-ph]} \BibitemShut
  {NoStop}%
\bibitem [{\citenamefont {Swanson}(2015)}]{Swanson:2014tra}%
  \BibitemOpen
  \bibfield  {author} {\bibinfo {author} {\bibfnamefont {E.~S.}\ \bibnamefont
  {Swanson}},\ }\bibfield  {title} {\bibinfo {title} {{$Z_b$ and $Z_c$ exotic
  states as coupled channel cusps}},\ }\href
  {https://doi.org/10.1103/PhysRevD.91.034009} {\bibfield  {journal} {\bibinfo
  {journal} {Phys. Rev. D}\ }\textbf {\bibinfo {volume} {91}},\ \bibinfo
  {pages} {034009} (\bibinfo {year} {2015})},\ \Eprint
  {https://arxiv.org/abs/1409.3291} {arXiv:1409.3291 [hep-ph]} \BibitemShut
  {NoStop}%
\bibitem [{\citenamefont {Albaladejo}\ \emph {et~al.}(2015)\citenamefont
  {Albaladejo}, \citenamefont {Guo}, \citenamefont {{Hidalgo-Duque}},
  \citenamefont {Nieves},\ and\ \citenamefont
  {Valderrama}}]{Albaladejo:2015dsa}%
  \BibitemOpen
  \bibfield  {author} {\bibinfo {author} {\bibfnamefont {M.}~\bibnamefont
  {Albaladejo}}, \bibinfo {author} {\bibfnamefont {F.-K.}\ \bibnamefont {Guo}},
  \bibinfo {author} {\bibfnamefont {C.}~\bibnamefont {{Hidalgo-Duque}}},
  \bibinfo {author} {\bibfnamefont {J.}~\bibnamefont {Nieves}},\ and\ \bibinfo
  {author} {\bibfnamefont {M.~P.}\ \bibnamefont {Valderrama}},\ }\bibfield
  {title} {\bibinfo {title} {{Decay widths of the spin-2 partners of the
  $X(3872)$}},\ }\href {https://doi.org/10.1140/epjc/s10052-015-3753-6}
  {\bibfield  {journal} {\bibinfo  {journal} {Eur. Phys. J. C}\ }\textbf
  {\bibinfo {volume} {75}},\ \bibinfo {pages} {547} (\bibinfo {year} {2015})},\
  \Eprint {https://arxiv.org/abs/1504.00861} {arXiv:1504.00861 [hep-ph]}
  \BibitemShut {NoStop}%
\bibitem [{\citenamefont {Albaladejo}\ \emph
  {et~al.}(2016{\natexlab{a}})\citenamefont {Albaladejo}, \citenamefont {Guo},
  \citenamefont {Hidalgo-Duque},\ and\ \citenamefont
  {Nieves}}]{Albaladejo:2015lob}%
  \BibitemOpen
  \bibfield  {author} {\bibinfo {author} {\bibfnamefont {M.}~\bibnamefont
  {Albaladejo}}, \bibinfo {author} {\bibfnamefont {F.-K.}\ \bibnamefont {Guo}},
  \bibinfo {author} {\bibfnamefont {C.}~\bibnamefont {Hidalgo-Duque}},\ and\
  \bibinfo {author} {\bibfnamefont {J.}~\bibnamefont {Nieves}},\ }\bibfield
  {title} {\bibinfo {title} {{$Z_c(3900)$: What has been really seen?}},\
  }\href {https://doi.org/10.1016/j.physletb.2016.02.025} {\bibfield  {journal}
  {\bibinfo  {journal} {Phys. Lett. B}\ }\textbf {\bibinfo {volume} {755}},\
  \bibinfo {pages} {337} (\bibinfo {year} {2016}{\natexlab{a}})},\ \Eprint
  {https://arxiv.org/abs/1512.03638} {arXiv:1512.03638 [hep-ph]} \BibitemShut
  {NoStop}%
\bibitem [{\citenamefont {Baru}\ \emph {et~al.}(2015)\citenamefont {Baru},
  \citenamefont {Epelbaum}, \citenamefont {Filin}, \citenamefont {Guo},
  \citenamefont {Hammer}, \citenamefont {Hanhart}, \citenamefont
  {Mei\ss{}ner},\ and\ \citenamefont {Nefediev}}]{Baru:2015nea}%
  \BibitemOpen
  \bibfield  {author} {\bibinfo {author} {\bibfnamefont {V.}~\bibnamefont
  {Baru}}, \bibinfo {author} {\bibfnamefont {E.}~\bibnamefont {Epelbaum}},
  \bibinfo {author} {\bibfnamefont {A.~A.}\ \bibnamefont {Filin}}, \bibinfo
  {author} {\bibfnamefont {F.-K.}\ \bibnamefont {Guo}}, \bibinfo {author}
  {\bibfnamefont {H.-W.}\ \bibnamefont {Hammer}}, \bibinfo {author}
  {\bibfnamefont {C.}~\bibnamefont {Hanhart}}, \bibinfo {author} {\bibfnamefont
  {U.-G.}\ \bibnamefont {Mei\ss{}ner}},\ and\ \bibinfo {author} {\bibfnamefont
  {A.~V.}\ \bibnamefont {Nefediev}},\ }\bibfield  {title} {\bibinfo {title}
  {{Remarks on study of $X(3872)$ from effective field theory with
  pion-exchange interaction}},\ }\href
  {https://doi.org/10.1103/PhysRevD.91.034002} {\bibfield  {journal} {\bibinfo
  {journal} {Phys. Rev. D}\ }\textbf {\bibinfo {volume} {91}},\ \bibinfo
  {pages} {034002} (\bibinfo {year} {2015})},\ \Eprint
  {https://arxiv.org/abs/1501.02924} {arXiv:1501.02924 [hep-ph]} \BibitemShut
  {NoStop}%
\bibitem [{\citenamefont {Szczepaniak}(2015)}]{Szczepaniak:2015eza}%
  \BibitemOpen
  \bibfield  {author} {\bibinfo {author} {\bibfnamefont {A.~P.}\ \bibnamefont
  {Szczepaniak}},\ }\bibfield  {title} {\bibinfo {title} {{Triangle
  singularities and $XYZ$ quarkonium peaks}},\ }\href
  {https://doi.org/10.1016/j.physletb.2015.06.029} {\bibfield  {journal}
  {\bibinfo  {journal} {Phys. Lett. B}\ }\textbf {\bibinfo {volume} {747}},\
  \bibinfo {pages} {410} (\bibinfo {year} {2015})},\ \Eprint
  {https://arxiv.org/abs/1501.01691} {arXiv:1501.01691 [hep-ph]} \BibitemShut
  {NoStop}%
\bibitem [{\citenamefont {Albaladejo}\ \emph
  {et~al.}(2016{\natexlab{b}})\citenamefont {Albaladejo}, \citenamefont
  {Fernandez-Soler},\ and\ \citenamefont {Nieves}}]{Albaladejo:2016jsg}%
  \BibitemOpen
  \bibfield  {author} {\bibinfo {author} {\bibfnamefont {M.}~\bibnamefont
  {Albaladejo}}, \bibinfo {author} {\bibfnamefont {P.}~\bibnamefont
  {Fernandez-Soler}},\ and\ \bibinfo {author} {\bibfnamefont {J.}~\bibnamefont
  {Nieves}},\ }\bibfield  {title} {\bibinfo {title} {{$Z_c(3900)$: Confronting
  theory and lattice simulations}},\ }\href
  {https://doi.org/10.1140/epjc/s10052-016-4427-8} {\bibfield  {journal}
  {\bibinfo  {journal} {Eur. Phys. J. C}\ }\textbf {\bibinfo {volume} {76}},\
  \bibinfo {pages} {573} (\bibinfo {year} {2016}{\natexlab{b}})},\ \Eprint
  {https://arxiv.org/abs/1606.03008} {arXiv:1606.03008 [hep-ph]} \BibitemShut
  {NoStop}%
\bibitem [{\citenamefont {Chen}\ \emph {et~al.}(2016)\citenamefont {Chen},
  \citenamefont {Chen}, \citenamefont {Liu},\ and\ \citenamefont
  {Zhu}}]{Chen:2016qju}%
  \BibitemOpen
  \bibfield  {author} {\bibinfo {author} {\bibfnamefont {H.-X.}\ \bibnamefont
  {Chen}}, \bibinfo {author} {\bibfnamefont {W.}~\bibnamefont {Chen}}, \bibinfo
  {author} {\bibfnamefont {X.}~\bibnamefont {Liu}},\ and\ \bibinfo {author}
  {\bibfnamefont {S.-L.}\ \bibnamefont {Zhu}},\ }\bibfield  {title} {\bibinfo
  {title} {{The hidden-charm pentaquark and tetraquark states}},\ }\href
  {https://doi.org/10.1016/j.physrep.2016.05.004} {\bibfield  {journal}
  {\bibinfo  {journal} {Phys. Rept.}\ }\textbf {\bibinfo {volume} {639}},\
  \bibinfo {pages} {1} (\bibinfo {year} {2016})},\ \Eprint
  {https://arxiv.org/abs/1601.02092} {arXiv:1601.02092 [hep-ph]} \BibitemShut
  {NoStop}%
\bibitem [{\citenamefont {Cincioglu}\ \emph {et~al.}(2016)\citenamefont
  {Cincioglu}, \citenamefont {Nieves}, \citenamefont {Ozpineci},\ and\
  \citenamefont {Yilmazer}}]{Cincioglu:2016fkm}%
  \BibitemOpen
  \bibfield  {author} {\bibinfo {author} {\bibfnamefont {E.}~\bibnamefont
  {Cincioglu}}, \bibinfo {author} {\bibfnamefont {J.}~\bibnamefont {Nieves}},
  \bibinfo {author} {\bibfnamefont {A.}~\bibnamefont {Ozpineci}},\ and\
  \bibinfo {author} {\bibfnamefont {A.~U.}\ \bibnamefont {Yilmazer}},\
  }\bibfield  {title} {\bibinfo {title} {{Quarkonium contribution to meson
  molecules}},\ }\href {https://doi.org/10.1140/epjc/s10052-016-4413-1}
  {\bibfield  {journal} {\bibinfo  {journal} {Eur. Phys. J. C}\ }\textbf
  {\bibinfo {volume} {76}},\ \bibinfo {pages} {576} (\bibinfo {year} {2016})},\
  \Eprint {https://arxiv.org/abs/1606.03239} {arXiv:1606.03239 [hep-ph]}
  \BibitemShut {NoStop}%
\bibitem [{\citenamefont {Albaladejo}\ \emph
  {et~al.}(2017{\natexlab{a}})\citenamefont {Albaladejo}, \citenamefont {Guo},
  \citenamefont {Hanhart}, \citenamefont {Mei\ss{}ner}, \citenamefont {Nieves},
  \citenamefont {Nogga},\ and\ \citenamefont {Yang}}]{Albaladejo:2017blx}%
  \BibitemOpen
  \bibfield  {author} {\bibinfo {author} {\bibfnamefont {M.}~\bibnamefont
  {Albaladejo}}, \bibinfo {author} {\bibfnamefont {F.-K.}\ \bibnamefont {Guo}},
  \bibinfo {author} {\bibfnamefont {C.}~\bibnamefont {Hanhart}}, \bibinfo
  {author} {\bibfnamefont {U.-G.}\ \bibnamefont {Mei\ss{}ner}}, \bibinfo
  {author} {\bibfnamefont {J.}~\bibnamefont {Nieves}}, \bibinfo {author}
  {\bibfnamefont {A.}~\bibnamefont {Nogga}},\ and\ \bibinfo {author}
  {\bibfnamefont {Z.}~\bibnamefont {Yang}},\ }\bibfield  {title} {\bibinfo
  {title} {{Note on $X(3872)$ production at hadron colliders and its molecular
  structure}},\ }\href {https://doi.org/10.1088/1674-1137/41/12/121001}
  {\bibfield  {journal} {\bibinfo  {journal} {Chin. Phys. C}\ }\textbf
  {\bibinfo {volume} {41}},\ \bibinfo {pages} {121001} (\bibinfo {year}
  {2017}{\natexlab{a}})},\ \Eprint {https://arxiv.org/abs/1709.09101}
  {arXiv:1709.09101 [hep-ph]} \BibitemShut {NoStop}%
\bibitem [{\citenamefont {Guo}\ \emph {et~al.}(2018)\citenamefont {Guo},
  \citenamefont {Hanhart}, \citenamefont {Mei{\ss}ner}, \citenamefont {Wang},
  \citenamefont {Zhao},\ and\ \citenamefont {Zou}}]{Guo:2017jvc}%
  \BibitemOpen
  \bibfield  {author} {\bibinfo {author} {\bibfnamefont {F.-K.}\ \bibnamefont
  {Guo}}, \bibinfo {author} {\bibfnamefont {C.}~\bibnamefont {Hanhart}},
  \bibinfo {author} {\bibfnamefont {U.-G.}\ \bibnamefont {Mei{\ss}ner}},
  \bibinfo {author} {\bibfnamefont {Q.}~\bibnamefont {Wang}}, \bibinfo {author}
  {\bibfnamefont {Q.}~\bibnamefont {Zhao}},\ and\ \bibinfo {author}
  {\bibfnamefont {B.-S.}\ \bibnamefont {Zou}},\ }\bibfield  {title} {\bibinfo
  {title} {Hadronic molecules},\ }\href
  {https://doi.org/10.1103/RevModPhys.90.015004} {\bibfield  {journal}
  {\bibinfo  {journal} {Rev. Mod. Phys.}\ }\textbf {\bibinfo {volume} {90}},\
  \bibinfo {pages} {015004} (\bibinfo {year} {2018})},\ \Eprint
  {https://arxiv.org/abs/1705.00141} {arXiv:1705.00141 [hep-ph]} \BibitemShut
  {NoStop}%
\bibitem [{\citenamefont {Guo}(2019)}]{Guo:2019qcn}%
  \BibitemOpen
  \bibfield  {author} {\bibinfo {author} {\bibfnamefont {F.-K.}\ \bibnamefont
  {Guo}},\ }\bibfield  {title} {\bibinfo {title} {{Novel Method for Precisely
  Measuring the $X(3872)$ Mass}},\ }\href
  {https://doi.org/10.1103/PhysRevLett.122.202002} {\bibfield  {journal}
  {\bibinfo  {journal} {Phys. Rev. Lett.}\ }\textbf {\bibinfo {volume} {122}},\
  \bibinfo {pages} {202002} (\bibinfo {year} {2019})},\ \Eprint
  {https://arxiv.org/abs/1902.11221} {arXiv:1902.11221 [hep-ph]} \BibitemShut
  {NoStop}%
\bibitem [{\citenamefont {Voloshin}(2019)}]{Voloshin:2019ivc}%
  \BibitemOpen
  \bibfield  {author} {\bibinfo {author} {\bibfnamefont {M.~B.}\ \bibnamefont
  {Voloshin}},\ }\bibfield  {title} {\bibinfo {title} {{Radiative and pionic
  transitions $Z_c(4020)^0 \to X(3872) \gamma$ and $Z_c(4020)^\pm \to X(3872)
  \pi^\pm$}},\ }\href {https://doi.org/10.1103/PhysRevD.99.054028} {\bibfield
  {journal} {\bibinfo  {journal} {Phys. Rev. D}\ }\textbf {\bibinfo {volume}
  {99}},\ \bibinfo {pages} {054028} (\bibinfo {year} {2019})},\ \Eprint
  {https://arxiv.org/abs/1902.01281} {arXiv:1902.01281 [hep-ph]} \BibitemShut
  {NoStop}%
\bibitem [{\citenamefont {Yang}\ \emph {et~al.}(2021)\citenamefont {Yang},
  \citenamefont {Cao}, \citenamefont {Guo}, \citenamefont {Nieves},\ and\
  \citenamefont {Valderrama}}]{Yang:2020nrt}%
  \BibitemOpen
  \bibfield  {author} {\bibinfo {author} {\bibfnamefont {Z.}~\bibnamefont
  {Yang}}, \bibinfo {author} {\bibfnamefont {X.}~\bibnamefont {Cao}}, \bibinfo
  {author} {\bibfnamefont {F.-K.}\ \bibnamefont {Guo}}, \bibinfo {author}
  {\bibfnamefont {J.}~\bibnamefont {Nieves}},\ and\ \bibinfo {author}
  {\bibfnamefont {M.~P.}\ \bibnamefont {Valderrama}},\ }\bibfield  {title}
  {\bibinfo {title} {Strange molecular partners of the
  {{Z}}{\textsubscript{c}}(3900) and {{Z}}{\textsubscript{c}}(4020)},\ }\href
  {https://doi.org/10.1103/PhysRevD.103.074029} {\bibfield  {journal} {\bibinfo
   {journal} {Phys. Rev. D}\ }\textbf {\bibinfo {volume} {103}},\ \bibinfo
  {pages} {074029} (\bibinfo {year} {2021})},\ \Eprint
  {https://arxiv.org/abs/2011.08725} {arXiv:2011.08725 [hep-ph]} \BibitemShut
  {NoStop}%
\bibitem [{\citenamefont {Zhang}\ and\ \citenamefont
  {Guo}(2021)}]{Zhang:2020mpi}%
  \BibitemOpen
  \bibfield  {author} {\bibinfo {author} {\bibfnamefont {Z.-H.}\ \bibnamefont
  {Zhang}}\ and\ \bibinfo {author} {\bibfnamefont {F.-K.}\ \bibnamefont
  {Guo}},\ }\bibfield  {title} {\bibinfo {title} {{$D^{\pm}D^{*\mp}$ hadronic
  atom as a key to revealing the $X(3872)$ mystery}},\ }\href
  {https://doi.org/10.1103/PhysRevLett.127.012002} {\bibfield  {journal}
  {\bibinfo  {journal} {Phys. Rev. Lett.}\ }\textbf {\bibinfo {volume} {127}},\
  \bibinfo {pages} {012002} (\bibinfo {year} {2021})},\ \Eprint
  {https://arxiv.org/abs/2012.08281} {arXiv:2012.08281 [hep-ph]} \BibitemShut
  {NoStop}%
\bibitem [{\citenamefont {Dong}\ \emph
  {et~al.}(2021{\natexlab{a}})\citenamefont {Dong}, \citenamefont {Guo},\ and\
  \citenamefont {Zou}}]{Dong:2020hxe}%
  \BibitemOpen
  \bibfield  {author} {\bibinfo {author} {\bibfnamefont {X.-K.}\ \bibnamefont
  {Dong}}, \bibinfo {author} {\bibfnamefont {F.-K.}\ \bibnamefont {Guo}},\ and\
  \bibinfo {author} {\bibfnamefont {B.-S.}\ \bibnamefont {Zou}},\ }\bibfield
  {title} {\bibinfo {title} {{Explaining the many threshold structures in the
  heavy-quark hadron spectrum}},\ }\href
  {https://doi.org/10.1103/PhysRevLett.126.152001} {\bibfield  {journal}
  {\bibinfo  {journal} {Phys. Rev. Lett.}\ }\textbf {\bibinfo {volume} {126}},\
  \bibinfo {pages} {152001} (\bibinfo {year} {2021}{\natexlab{a}})},\ \Eprint
  {https://arxiv.org/abs/2011.14517} {arXiv:2011.14517 [hep-ph]} \BibitemShut
  {NoStop}%
\bibitem [{\citenamefont {Brambilla}\ \emph {et~al.}(2020)\citenamefont
  {Brambilla}, \citenamefont {Eidelman}, \citenamefont {Hanhart}, \citenamefont
  {Nefediev}, \citenamefont {Shen}, \citenamefont {Thomas}, \citenamefont
  {Vairo},\ and\ \citenamefont {Yuan}}]{Brambilla:2019esw}%
  \BibitemOpen
  \bibfield  {author} {\bibinfo {author} {\bibfnamefont {N.}~\bibnamefont
  {Brambilla}}, \bibinfo {author} {\bibfnamefont {S.}~\bibnamefont {Eidelman}},
  \bibinfo {author} {\bibfnamefont {C.}~\bibnamefont {Hanhart}}, \bibinfo
  {author} {\bibfnamefont {A.}~\bibnamefont {Nefediev}}, \bibinfo {author}
  {\bibfnamefont {C.-P.}\ \bibnamefont {Shen}}, \bibinfo {author}
  {\bibfnamefont {C.~E.}\ \bibnamefont {Thomas}}, \bibinfo {author}
  {\bibfnamefont {A.}~\bibnamefont {Vairo}},\ and\ \bibinfo {author}
  {\bibfnamefont {C.-Z.}\ \bibnamefont {Yuan}},\ }\bibfield  {title} {\bibinfo
  {title} {The {{$XYZ$}} states: {{Experimental}} and theoretical status and
  perspectives},\ }\href {https://doi.org/10.1016/j.physrep.2020.05.001}
  {\bibfield  {journal} {\bibinfo  {journal} {Phys. Rept.}\ }\textbf {\bibinfo
  {volume} {873}},\ \bibinfo {pages} {1} (\bibinfo {year} {2020})},\ \Eprint
  {https://arxiv.org/abs/1907.07583} {arXiv:1907.07583 [hep-ex]} \BibitemShut
  {NoStop}%
\bibitem [{\citenamefont {Dong}\ \emph
  {et~al.}(2021{\natexlab{b}})\citenamefont {Dong}, \citenamefont {Guo},\ and\
  \citenamefont {Zou}}]{Dong:2021bvy}%
  \BibitemOpen
  \bibfield  {author} {\bibinfo {author} {\bibfnamefont {X.-K.}\ \bibnamefont
  {Dong}}, \bibinfo {author} {\bibfnamefont {F.-K.}\ \bibnamefont {Guo}},\ and\
  \bibinfo {author} {\bibfnamefont {B.-S.}\ \bibnamefont {Zou}},\ }\bibfield
  {title} {\bibinfo {title} {{A survey of heavy\textendash{}heavy hadronic
  molecules}},\ }\href {https://doi.org/10.1088/1572-9494/ac27a2} {\bibfield
  {journal} {\bibinfo  {journal} {Commun. Theor. Phys.}\ }\textbf {\bibinfo
  {volume} {73}},\ \bibinfo {pages} {125201} (\bibinfo {year}
  {2021}{\natexlab{b}})},\ \Eprint {https://arxiv.org/abs/2108.02673}
  {arXiv:2108.02673 [hep-ph]} \BibitemShut {NoStop}%
\bibitem [{\citenamefont {Baru}\ \emph {et~al.}(2022)\citenamefont {Baru},
  \citenamefont {Epelbaum}, \citenamefont {Filin}, \citenamefont {Hanhart},\
  and\ \citenamefont {Nefediev}}]{Baru:2021ddn}%
  \BibitemOpen
  \bibfield  {author} {\bibinfo {author} {\bibfnamefont {V.}~\bibnamefont
  {Baru}}, \bibinfo {author} {\bibfnamefont {E.}~\bibnamefont {Epelbaum}},
  \bibinfo {author} {\bibfnamefont {A.~A.}\ \bibnamefont {Filin}}, \bibinfo
  {author} {\bibfnamefont {C.}~\bibnamefont {Hanhart}},\ and\ \bibinfo {author}
  {\bibfnamefont {A.~V.}\ \bibnamefont {Nefediev}},\ }\bibfield  {title}
  {\bibinfo {title} {{Is $Z_{cs}(3982)$ a molecular partner of $Z_c(3900)$ and
  $Z_c(4020)$ states?}},\ }\href {https://doi.org/10.1103/PhysRevD.105.034014}
  {\bibfield  {journal} {\bibinfo  {journal} {Phys. Rev. D}\ }\textbf {\bibinfo
  {volume} {105}},\ \bibinfo {pages} {034014} (\bibinfo {year} {2022})},\
  \Eprint {https://arxiv.org/abs/2110.00398} {arXiv:2110.00398 [hep-ph]}
  \BibitemShut {NoStop}%
\bibitem [{\citenamefont {Du}\ \emph {et~al.}(2022)\citenamefont {Du},
  \citenamefont {Albaladejo}, \citenamefont {Guo},\ and\ \citenamefont
  {Nieves}}]{Du:2022jjv}%
  \BibitemOpen
  \bibfield  {author} {\bibinfo {author} {\bibfnamefont {M.-L.}\ \bibnamefont
  {Du}}, \bibinfo {author} {\bibfnamefont {M.}~\bibnamefont {Albaladejo}},
  \bibinfo {author} {\bibfnamefont {F.-K.}\ \bibnamefont {Guo}},\ and\ \bibinfo
  {author} {\bibfnamefont {J.}~\bibnamefont {Nieves}},\ }\bibfield  {title}
  {\bibinfo {title} {{Combined analysis of the $Z_c(3900)$ and the
  $Z_{cs}(3985)$ exotic states}},\ }\href
  {https://doi.org/10.1103/PhysRevD.105.074018} {\bibfield  {journal} {\bibinfo
   {journal} {Phys. Rev. D}\ }\textbf {\bibinfo {volume} {105}},\ \bibinfo
  {pages} {074018} (\bibinfo {year} {2022})},\ \Eprint
  {https://arxiv.org/abs/2201.08253} {arXiv:2201.08253 [hep-ph]} \BibitemShut
  {NoStop}%
\bibitem [{\citenamefont {Ablikim}\ \emph
  {et~al.}(2013{\natexlab{b}})\citenamefont {Ablikim} \emph
  {et~al.}}]{BESIII:2013ouc}%
  \BibitemOpen
  \bibfield  {author} {\bibinfo {author} {\bibfnamefont {M.}~\bibnamefont
  {Ablikim}} \emph {et~al.} (\bibinfo {collaboration} {BESIII}),\ }\bibfield
  {title} {\bibinfo {title} {Observation of a charged charmoniumlike structure
  {{$Z_c(4020)$ and search for the $Z_c(3900)$}} in $e^+e^-\to
  \pi^{+}\pi^-h_c$},\ }\href {https://doi.org/10.1103/PhysRevLett.111.242001}
  {\bibfield  {journal} {\bibinfo  {journal} {Phys. Rev. Lett.}\ }\textbf
  {\bibinfo {volume} {111}},\ \bibinfo {pages} {242001} (\bibinfo {year}
  {2013}{\natexlab{b}})},\ \Eprint {https://arxiv.org/abs/1309.1896}
  {arXiv:1309.1896 [hep-ex]} \BibitemShut {NoStop}%
\bibitem [{\citenamefont {Ablikim}\ \emph {et~al.}(2021)\citenamefont {Ablikim}
  \emph {et~al.}}]{BESIII:2020qkh}%
  \BibitemOpen
  \bibfield  {author} {\bibinfo {author} {\bibfnamefont {M.}~\bibnamefont
  {Ablikim}} \emph {et~al.} (\bibinfo {collaboration} {BESIII}),\ }\bibfield
  {title} {\bibinfo {title} {{Observation of a Near-Threshold Structure in the
  $K^+$ Recoil-Mass Spectra in $e^+e^- \rightarrow
  K^+(D_s^-D^{*0}+D_s^{*-}D^0$)}},\ }\href
  {https://doi.org/10.1103/PhysRevLett.126.102001} {\bibfield  {journal}
  {\bibinfo  {journal} {Phys. Rev. Lett.}\ }\textbf {\bibinfo {volume} {126}},\
  \bibinfo {pages} {102001} (\bibinfo {year} {2021})},\ \Eprint
  {https://arxiv.org/abs/2011.07855} {arXiv:2011.07855 [hep-ex]} \BibitemShut
  {NoStop}%
\bibitem [{\citenamefont {Aaij}\ \emph {et~al.}(2021)\citenamefont {Aaij} \emph
  {et~al.}}]{LHCb:2021uow}%
  \BibitemOpen
  \bibfield  {author} {\bibinfo {author} {\bibfnamefont {R.}~\bibnamefont
  {Aaij}} \emph {et~al.} (\bibinfo {collaboration} {LHCb}),\ }\bibfield
  {title} {\bibinfo {title} {Observation of new resonances decaying to
  {{$J/\psi K^+$}} and {{$J/\psi\phi$}}},\ }\href
  {https://doi.org/10.1103/PhysRevLett.127.082001} {\bibfield  {journal}
  {\bibinfo  {journal} {Phys. Rev. Lett.}\ }\textbf {\bibinfo {volume} {127}},\
  \bibinfo {pages} {082001} (\bibinfo {year} {2021})},\ \Eprint
  {https://arxiv.org/abs/2103.01803} {arXiv:2103.01803 [hep-ex]} \BibitemShut
  {NoStop}%
\bibitem [{\citenamefont {Spadaro~Norella}\ and\ \citenamefont
  {Chen}(2022)}]{LHCb:2022NewObservations}%
  \BibitemOpen
  \bibfield  {author} {\bibinfo {author} {\bibfnamefont {E.}~\bibnamefont
  {Spadaro~Norella}}\ and\ \bibinfo {author} {\bibfnamefont {C.}~\bibnamefont
  {Chen}} (\bibinfo {collaboration} {LHCb}),\ }\href
  {https://indico.cern.ch/event/1176505/} {\bibinfo {title} {Particle {{Zoo}}
  2.0: New {{Tetra}}- and {{Pentaquarks}} at {{LHCb}}: {\tt
  https://indico.cern.ch/event/1176505/}}} (\bibinfo {year} {2022})\BibitemShut
  {NoStop}%
\bibitem [{\citenamefont {Li}\ and\ \citenamefont
  {Voloshin}(2015)}]{Li:2015iga}%
  \BibitemOpen
  \bibfield  {author} {\bibinfo {author} {\bibfnamefont {X.}~\bibnamefont
  {Li}}\ and\ \bibinfo {author} {\bibfnamefont {M.~B.}\ \bibnamefont
  {Voloshin}},\ }\bibfield  {title} {\bibinfo {title} {{$X(3915)$ as a $D_s
  \bar D_s$ bound state}},\ }\href {https://doi.org/10.1103/PhysRevD.91.114014}
  {\bibfield  {journal} {\bibinfo  {journal} {Phys. Rev. D}\ }\textbf {\bibinfo
  {volume} {91}},\ \bibinfo {pages} {114014} (\bibinfo {year} {2015})},\
  \Eprint {https://arxiv.org/abs/1503.04431} {arXiv:1503.04431 [hep-ph]}
  \BibitemShut {NoStop}%
\bibitem [{\citenamefont {Prelovsek}\ \emph {et~al.}(2021)\citenamefont
  {Prelovsek}, \citenamefont {Collins}, \citenamefont {Mohler}, \citenamefont
  {Padmanath},\ and\ \citenamefont {Piemonte}}]{Prelovsek:2020eiw}%
  \BibitemOpen
  \bibfield  {author} {\bibinfo {author} {\bibfnamefont {S.}~\bibnamefont
  {Prelovsek}}, \bibinfo {author} {\bibfnamefont {S.}~\bibnamefont {Collins}},
  \bibinfo {author} {\bibfnamefont {D.}~\bibnamefont {Mohler}}, \bibinfo
  {author} {\bibfnamefont {M.}~\bibnamefont {Padmanath}},\ and\ \bibinfo
  {author} {\bibfnamefont {S.}~\bibnamefont {Piemonte}},\ }\bibfield  {title}
  {\bibinfo {title} {{Charmonium-like resonances with $J^{PC}$ = 0$^{++}$,
  2$^{++}$ in coupled $ {D}\overline{{D}} $, $
  {{D}}_{{s}}{\overline{{D}}}_{{s}} $ scattering on the lattice}},\ }\href
  {https://doi.org/10.1007/JHEP06(2021)035} {\bibfield  {journal} {\bibinfo
  {journal} {JHEP}\ }\textbf {\bibinfo {volume} {06}},\ \bibinfo {pages}
  {035}},\ \Eprint {https://arxiv.org/abs/2011.02542} {arXiv:2011.02542
  [hep-lat]} \BibitemShut {NoStop}%
\bibitem [{\citenamefont {Meng}\ \emph {et~al.}(2021)\citenamefont {Meng},
  \citenamefont {Wang},\ and\ \citenamefont {Zhu}}]{Meng:2020cbk}%
  \BibitemOpen
  \bibfield  {author} {\bibinfo {author} {\bibfnamefont {L.}~\bibnamefont
  {Meng}}, \bibinfo {author} {\bibfnamefont {B.}~\bibnamefont {Wang}},\ and\
  \bibinfo {author} {\bibfnamefont {S.-L.}\ \bibnamefont {Zhu}},\ }\bibfield
  {title} {\bibinfo {title} {{Predicting the $\bar D_s^{(*)}D_s^{(*)}$ bound
  states as the partners of $X(3872)$}},\ }\href
  {https://doi.org/10.1016/j.scib.2021.03.016} {\bibfield  {journal} {\bibinfo
  {journal} {Sci. Bull.}\ }\textbf {\bibinfo {volume} {66}},\ \bibinfo {pages}
  {1288} (\bibinfo {year} {2021})},\ \Eprint {https://arxiv.org/abs/2012.09813}
  {arXiv:2012.09813 [hep-ph]} \BibitemShut {NoStop}%
\bibitem [{\citenamefont {Dong}\ \emph
  {et~al.}(2021{\natexlab{c}})\citenamefont {Dong}, \citenamefont {Guo},\ and\
  \citenamefont {Zou}}]{Dong:2021juy}%
  \BibitemOpen
  \bibfield  {author} {\bibinfo {author} {\bibfnamefont {X.-K.}\ \bibnamefont
  {Dong}}, \bibinfo {author} {\bibfnamefont {F.-K.}\ \bibnamefont {Guo}},\ and\
  \bibinfo {author} {\bibfnamefont {B.-S.}\ \bibnamefont {Zou}},\ }\bibfield
  {title} {\bibinfo {title} {A survey of heavy-antiheavy hadronic molecules},\
  }\href {https://doi.org/10.13725/j.cnki.pip.2021.02.001} {\bibfield
  {journal} {\bibinfo  {journal} {Progr. Phys.}\ }\textbf {\bibinfo {volume}
  {41}},\ \bibinfo {pages} {65} (\bibinfo {year} {2021}{\natexlab{c}})},\
  \Eprint {https://arxiv.org/abs/2101.01021} {arXiv:2101.01021 [hep-ph]}
  \BibitemShut {NoStop}%
\bibitem [{\citenamefont {Aaij}\ \emph {et~al.}(2020)\citenamefont {Aaij} \emph
  {et~al.}}]{LHCb:2020xds}%
  \BibitemOpen
  \bibfield  {author} {\bibinfo {author} {\bibfnamefont {R.}~\bibnamefont
  {Aaij}} \emph {et~al.} (\bibinfo {collaboration} {LHCb}),\ }\bibfield
  {title} {\bibinfo {title} {Study of the lineshape of the
  {{$\chi_{c1}(3872)$}} state},\ }\href
  {https://doi.org/10.1103/PhysRevD.102.092005} {\bibfield  {journal} {\bibinfo
   {journal} {Phys. Rev. D}\ }\textbf {\bibinfo {volume} {102}},\ \bibinfo
  {pages} {092005} (\bibinfo {year} {2020})},\ \Eprint
  {https://arxiv.org/abs/2005.13419} {arXiv:2005.13419 [hep-ex]} \BibitemShut
  {NoStop}%
\bibitem [{\citenamefont {Aaij}\ and\ \citenamefont
  {{others}}(2022)}]{LHCb:2022bly}%
  \BibitemOpen
  \bibfield  {author} {\bibinfo {author} {\bibfnamefont {R.}~\bibnamefont
  {Aaij}}\ and\ \bibinfo {author} {\bibnamefont {{others}}} (\bibinfo
  {collaboration} {LHCb}),\ }\bibfield  {title} {\bibinfo {title} {Observation
  of sizeable $\omega$ contribution to $\chi_{c1}(3872)\to\pi^+\pi^-{{J}}/\psi$
  decays},\ }\href {https://arxiv.org/abs/2204.12597} {\bibfield  {journal}
  {\bibinfo  {journal} {arXiv:2204.12597 [hep-ex]}\ } (\bibinfo {year}
  {2022})},\ \Eprint {https://arxiv.org/abs/2204.12597} {arXiv:2204.12597
  [hep-ex]} \BibitemShut {NoStop}%
\bibitem [{\citenamefont {Ablikim}\ \emph {et~al.}(2015)\citenamefont {Ablikim}
  \emph {et~al.}}]{BESIII:2015pqw}%
  \BibitemOpen
  \bibfield  {author} {\bibinfo {author} {\bibfnamefont {M.}~\bibnamefont
  {Ablikim}} \emph {et~al.} (\bibinfo {collaboration} {BESIII}),\ }\bibfield
  {title} {\bibinfo {title} {{Confirmation of a charged charmoniumlike state
  $Z_c(3885)^{\mp}$ in $e^+e^-\to\pi^{\pm}(D\bar{D}^*)^\mp$ with double $D$
  tag}},\ }\href {https://doi.org/10.1103/PhysRevD.92.092006} {\bibfield
  {journal} {\bibinfo  {journal} {Phys. Rev. D}\ }\textbf {\bibinfo {volume}
  {92}},\ \bibinfo {pages} {092006} (\bibinfo {year} {2015})},\ \Eprint
  {https://arxiv.org/abs/1509.01398} {arXiv:1509.01398 [hep-ex]} \BibitemShut
  {NoStop}%
\bibitem [{\citenamefont {Ablikim}\ \emph {et~al.}(2017)\citenamefont {Ablikim}
  \emph {et~al.}}]{BESIII:2017bua}%
  \BibitemOpen
  \bibfield  {author} {\bibinfo {author} {\bibfnamefont {M.}~\bibnamefont
  {Ablikim}} \emph {et~al.} (\bibinfo {collaboration} {BESIII}),\ }\bibfield
  {title} {\bibinfo {title} {{Determination of the spin and parity of the
  $Z_c(3900)$}},\ }\href {https://doi.org/10.1103/PhysRevLett.119.072001}
  {\bibfield  {journal} {\bibinfo  {journal} {Phys. Rev. Lett.}\ }\textbf
  {\bibinfo {volume} {119}},\ \bibinfo {pages} {072001} (\bibinfo {year}
  {2017})},\ \Eprint {https://arxiv.org/abs/1706.04100} {arXiv:1706.04100
  [hep-ex]} \BibitemShut {NoStop}%
\bibitem [{\citenamefont {Mehen}\ and\ \citenamefont
  {Powell}(2011)}]{Mehen:2011yh}%
  \BibitemOpen
  \bibfield  {author} {\bibinfo {author} {\bibfnamefont {T.}~\bibnamefont
  {Mehen}}\ and\ \bibinfo {author} {\bibfnamefont {J.~W.}\ \bibnamefont
  {Powell}},\ }\bibfield  {title} {\bibinfo {title} {Heavy quark symmetry
  predictions for weakly bound {{$B$-meson}} molecules},\ }\href
  {https://doi.org/10.1103/PhysRevD.84.114013} {\bibfield  {journal} {\bibinfo
  {journal} {Phys. Rev. D}\ }\textbf {\bibinfo {volume} {84}},\ \bibinfo
  {pages} {114013} (\bibinfo {year} {2011})},\ \Eprint
  {https://arxiv.org/abs/1109.3479} {arXiv:1109.3479 [hep-ph]} \BibitemShut
  {NoStop}%
\bibitem [{\citenamefont {Grinstein}\ \emph {et~al.}(1992)\citenamefont
  {Grinstein}, \citenamefont {Jenkins}, \citenamefont {Manohar}, \citenamefont
  {Savage},\ and\ \citenamefont {Wise}}]{Grinstein:1992qt}%
  \BibitemOpen
  \bibfield  {author} {\bibinfo {author} {\bibfnamefont {B.}~\bibnamefont
  {Grinstein}}, \bibinfo {author} {\bibfnamefont {E.~E.}\ \bibnamefont
  {Jenkins}}, \bibinfo {author} {\bibfnamefont {A.~V.}\ \bibnamefont
  {Manohar}}, \bibinfo {author} {\bibfnamefont {M.~J.}\ \bibnamefont
  {Savage}},\ and\ \bibinfo {author} {\bibfnamefont {M.~B.}\ \bibnamefont
  {Wise}},\ }\bibfield  {title} {\bibinfo {title} {{Chiral perturbation theory
  for $f_{\mathrm{D}_{\mathrm{s}}} / f_{\mathrm{D}}$ and
  $B_{\mathrm{B}_{\mathrm{s}}} / B_{\mathrm{B}}$}},\ }\href
  {https://doi.org/10.1016/0550-3213(92)90248-A} {\bibfield  {journal}
  {\bibinfo  {journal} {Nucl. Phys. B}\ }\textbf {\bibinfo {volume} {380}},\
  \bibinfo {pages} {369} (\bibinfo {year} {1992})},\ \Eprint
  {https://arxiv.org/abs/hep-ph/9204207} {arXiv:hep-ph/9204207} \BibitemShut
  {NoStop}%
\bibitem [{\citenamefont {Manohar}\ and\ \citenamefont
  {Wise}(2000)}]{Manohar:2000dt}%
  \BibitemOpen
  \bibfield  {author} {\bibinfo {author} {\bibfnamefont {A.~V.}\ \bibnamefont
  {Manohar}}\ and\ \bibinfo {author} {\bibfnamefont {M.~B.}\ \bibnamefont
  {Wise}},\ }\href@noop {} {\emph {\bibinfo {title} {{Heavy Quark Physics}}}}\
  (\bibinfo  {publisher} {Cambridge University Press},\ \bibinfo {year}
  {2000})\BibitemShut {NoStop}%
\bibitem [{\citenamefont {Guo}\ \emph {et~al.}(2006)\citenamefont {Guo},
  \citenamefont {Shen}, \citenamefont {Chiang}, \citenamefont {Ping},\ and\
  \citenamefont {Zou}}]{Guo:2006fu}%
  \BibitemOpen
  \bibfield  {author} {\bibinfo {author} {\bibfnamefont {F.-K.}\ \bibnamefont
  {Guo}}, \bibinfo {author} {\bibfnamefont {P.-N.}\ \bibnamefont {Shen}},
  \bibinfo {author} {\bibfnamefont {H.-C.}\ \bibnamefont {Chiang}}, \bibinfo
  {author} {\bibfnamefont {R.-G.}\ \bibnamefont {Ping}},\ and\ \bibinfo
  {author} {\bibfnamefont {B.-S.}\ \bibnamefont {Zou}},\ }\bibfield  {title}
  {\bibinfo {title} {{Dynamically generated 0+ heavy mesons in a heavy chiral
  unitary approach}},\ }\href {https://doi.org/10.1016/j.physletb.2006.08.064}
  {\bibfield  {journal} {\bibinfo  {journal} {Phys. Lett. B}\ }\textbf
  {\bibinfo {volume} {641}},\ \bibinfo {pages} {278} (\bibinfo {year}
  {2006})},\ \Eprint {https://arxiv.org/abs/hep-ph/0603072}
  {arXiv:hep-ph/0603072} \BibitemShut {NoStop}%
\bibitem [{\citenamefont {Albaladejo}\ \emph
  {et~al.}(2017{\natexlab{b}})\citenamefont {Albaladejo}, \citenamefont
  {Fernandez-Soler}, \citenamefont {Guo},\ and\ \citenamefont
  {Nieves}}]{Albaladejo:2016lbb}%
  \BibitemOpen
  \bibfield  {author} {\bibinfo {author} {\bibfnamefont {M.}~\bibnamefont
  {Albaladejo}}, \bibinfo {author} {\bibfnamefont {P.}~\bibnamefont
  {Fernandez-Soler}}, \bibinfo {author} {\bibfnamefont {F.-K.}\ \bibnamefont
  {Guo}},\ and\ \bibinfo {author} {\bibfnamefont {J.}~\bibnamefont {Nieves}},\
  }\bibfield  {title} {\bibinfo {title} {{Two-pole structure of the
  $D^\ast_0(2400)$}},\ }\href {https://doi.org/10.1016/j.physletb.2017.02.036}
  {\bibfield  {journal} {\bibinfo  {journal} {Phys. Lett. B}\ }\textbf
  {\bibinfo {volume} {767}},\ \bibinfo {pages} {465} (\bibinfo {year}
  {2017}{\natexlab{b}})},\ \Eprint {https://arxiv.org/abs/1610.06727}
  {arXiv:1610.06727 [hep-ph]} \BibitemShut {NoStop}%
\bibitem [{\citenamefont {Du}\ \emph {et~al.}(2018)\citenamefont {Du},
  \citenamefont {Albaladejo}, \citenamefont {Fern\'andez-Soler}, \citenamefont
  {Guo}, \citenamefont {Hanhart}, \citenamefont {Mei\ss{}ner}, \citenamefont
  {Nieves},\ and\ \citenamefont {Yao}}]{Du:2017zvv}%
  \BibitemOpen
  \bibfield  {author} {\bibinfo {author} {\bibfnamefont {M.-L.}\ \bibnamefont
  {Du}}, \bibinfo {author} {\bibfnamefont {M.}~\bibnamefont {Albaladejo}},
  \bibinfo {author} {\bibfnamefont {P.}~\bibnamefont {Fern\'andez-Soler}},
  \bibinfo {author} {\bibfnamefont {F.-K.}\ \bibnamefont {Guo}}, \bibinfo
  {author} {\bibfnamefont {C.}~\bibnamefont {Hanhart}}, \bibinfo {author}
  {\bibfnamefont {U.-G.}\ \bibnamefont {Mei\ss{}ner}}, \bibinfo {author}
  {\bibfnamefont {J.}~\bibnamefont {Nieves}},\ and\ \bibinfo {author}
  {\bibfnamefont {D.-L.}\ \bibnamefont {Yao}},\ }\bibfield  {title} {\bibinfo
  {title} {{Towards a new paradigm for heavy-light meson spectroscopy}},\
  }\href {https://doi.org/10.1103/PhysRevD.98.094018} {\bibfield  {journal}
  {\bibinfo  {journal} {Phys. Rev. D}\ }\textbf {\bibinfo {volume} {98}},\
  \bibinfo {pages} {094018} (\bibinfo {year} {2018})},\ \Eprint
  {https://arxiv.org/abs/1712.07957} {arXiv:1712.07957 [hep-ph]} \BibitemShut
  {NoStop}%
\bibitem [{\citenamefont {Aaij}\ \emph {et~al.}(2016)\citenamefont {Aaij} \emph
  {et~al.}}]{LHCb:2016lxy}%
  \BibitemOpen
  \bibfield  {author} {\bibinfo {author} {\bibfnamefont {R.}~\bibnamefont
  {Aaij}} \emph {et~al.} (\bibinfo {collaboration} {LHCb}),\ }\bibfield
  {title} {\bibinfo {title} {{Amplitude analysis of $B^{-} \to D^{+} \pi^{-}
  \pi^{-}$ decays}},\ }\href {https://doi.org/10.1103/PhysRevD.94.072001}
  {\bibfield  {journal} {\bibinfo  {journal} {Phys. Rev. D}\ }\textbf {\bibinfo
  {volume} {94}},\ \bibinfo {pages} {072001} (\bibinfo {year} {2016})},\
  \Eprint {https://arxiv.org/abs/1608.01289} {arXiv:1608.01289 [hep-ex]}
  \BibitemShut {NoStop}%
\bibitem [{\citenamefont {Workman}\ \emph {et~al.}(2022)\citenamefont {Workman}
  \emph {et~al.}}]{Workman:2022ynf}%
  \BibitemOpen
  \bibfield  {author} {\bibinfo {author} {\bibfnamefont {R.~L.}\ \bibnamefont
  {Workman}} \emph {et~al.} (\bibinfo {collaboration} {Particle Data Group}),\
  }\bibfield  {title} {\bibinfo {title} {Review of {{Particle Physics}}},\
  }\href@noop {} {\bibfield  {journal} {\bibinfo  {journal} {PTEP}\ }\textbf
  {\bibinfo {volume} {2022}},\ \bibinfo {pages} {083C01} (\bibinfo {year}
  {2022})}\BibitemShut {NoStop}%
\bibitem [{\citenamefont {Du}\ \emph {et~al.}(2021)\citenamefont {Du},
  \citenamefont {Baru}, \citenamefont {Guo}, \citenamefont {Hanhart},
  \citenamefont {Mei\ss{}ner}, \citenamefont {Oller},\ and\ \citenamefont
  {Wang}}]{Du:2021fmf}%
  \BibitemOpen
  \bibfield  {author} {\bibinfo {author} {\bibfnamefont {M.-L.}\ \bibnamefont
  {Du}}, \bibinfo {author} {\bibfnamefont {V.}~\bibnamefont {Baru}}, \bibinfo
  {author} {\bibfnamefont {F.-K.}\ \bibnamefont {Guo}}, \bibinfo {author}
  {\bibfnamefont {C.}~\bibnamefont {Hanhart}}, \bibinfo {author} {\bibfnamefont
  {U.-G.}\ \bibnamefont {Mei\ss{}ner}}, \bibinfo {author} {\bibfnamefont
  {J.~A.}\ \bibnamefont {Oller}},\ and\ \bibinfo {author} {\bibfnamefont
  {Q.}~\bibnamefont {Wang}},\ }\bibfield  {title} {\bibinfo {title}
  {{Revisiting the nature of the $P_{c}$ pentaquarks}},\ }\href
  {https://doi.org/10.1007/JHEP08(2021)157} {\bibfield  {journal} {\bibinfo
  {journal} {JHEP}\ }\textbf {\bibinfo {volume} {08}},\ \bibinfo {pages}
  {157}},\ \Eprint {https://arxiv.org/abs/2102.07159} {arXiv:2102.07159
  [hep-ph]} \BibitemShut {NoStop}%
\bibitem [{\citenamefont {Hanhart}\ and\ \citenamefont
  {Nefediev}(2022)}]{Hanhart:2022qxq}%
  \BibitemOpen
  \bibfield  {author} {\bibinfo {author} {\bibfnamefont {C.}~\bibnamefont
  {Hanhart}}\ and\ \bibinfo {author} {\bibfnamefont {A.~V.}\ \bibnamefont
  {Nefediev}},\ }\bibfield  {title} {\bibinfo {title} {{Do near threshold
  molecular states mix with neighbouring $\bar QQ$ states?}},\ }\href@noop {}
  {\  (\bibinfo {year} {2022})},\ \Eprint {https://arxiv.org/abs/2209.10165}
  {arXiv:2209.10165 [hep-ph]} \BibitemShut {NoStop}%
\end{thebibliography}%
 
\end{document}